\documentclass[a4paper,12pt]{article}

\usepackage{amssymb,amsmath,amsfonts,mathrsfs,mathbbol}
\usepackage{fullpage}

\usepackage{graphicx}
\usepackage{caption}
\def\ben{\begin{equation}}
\def\een{\end{equation}}
\def\bena{\begin{eqnarray}}
\def\eena{\end{eqnarray}}
\def\non{\nonumber}
\newcommand{\Sum}{\operatorname*{\sum^\prime}}

\newcommand{\Exp}{\mathrm{e}}

\newcommand{\slim}{\operatorname*{s-lim}}
\newcommand{\tr}{\mathrm{Tr}}
\newcommand{\e}{\mathrm{e}}

\newcommand{\mc}{\mathbb{C}}
\newcommand{\mr}{\mathbb{R}}

\newcommand{\half}{\frac{1}{2}}
\newcommand{\phalf}{{\frac{p}{2}}}

\newcommand{\QQ}{\mathcal{Q}}
\renewcommand{\P}{\mathrm{P}}
\newcommand{\Q}{\mathrm{Q}}
\renewcommand{\H}{\mathcal{H}}
\newcommand{\E}{\mathrm{E}}
\newcommand{\F}{\mathcal{F}}

\newcommand{\re}{\mathrm{Re}}
\newcommand{\im}{\mathrm{Im}}
\newcommand{\Int}{\int\limits}
\newcommand{\negfive}{\!\!\!\!\!}

\newcommand{\R}{\mathcal{R}}
\newcommand{\mz}{\mathbb{Z}}
\newcommand{\supp}{\operatorname*{supp}}
\newcommand{\bv}{\operatorname*{B.V.}}

\renewcommand{\c}{K}

\newcommand{\M}{\mathcal{M}}
\newcommand{\B}{\mathcal{B}}
\newcommand{\D}{\mathscr{D}}
\newcommand{\qed}{\hfill{$\square$}}

\newtheorem{thm}{Theorem}

\newtheorem{lemma}{Lemma}

\newtheorem{prop}{Proposition}

\title{On the derivation of the Boltzmann equation in quantum field theory: Flat spacetime}

\author{
Stefan Hollands$^{1,2}$\thanks{\tt HollandsS@Cardiff.ac.uk}\:
and
Gregor Leiler$^{1}$\thanks{\tt LeilerG@Cardiff.ac.uk}\:
\\ \\
{\it ${}^{1}$School of Mathematics, Cardiff University} \\
{\it Cardiff, United Kingdom} \medskip \\
{\it ${}^{2}$KEK Theory Center,
Institute of Particle and Nuclear Studies
} \\
{\it High Energy Accelerator Research Organization (KEK)
} \\
{\it Tsukuba, Japan} \\
}

\begin{document}

\date{8 March, 2010}

\maketitle

\tableofcontents

\pagebreak

\begin{abstract}
In this paper, we analyze in a mathematically rigorous fashion the validity of the Boltzmann
transport equation within quantum field theory. We work within the specific model of a
hermitian, scalar field with polynomial self-interaction in two-dimensional Minkowski space.
Our main results are as follows: Firstly, that one can obtain a non-perturbative, exact integro-differential
equation for the number densities, which we called the pre-Boltzmann equation. We secondly take the long-time-dilute-medium
limit of this equation, to obtain a simpler equation. This limiting equation is qualitatively similar to the Boltzmann equation, but it involves additional re-scattering terms. These terms disappear if perform a perturbation expansion in the coupling constant and ignore the loop corrections (Born approximation). If loop corrections are included, then we argue that for
consistency, one must also keep corresponding rescattering terms which are normally ignored.
Our analysis is hence of potential relevance
for physical applications of the Boltzmann equation wherein loop effects are essential, such as in the standard
scenario of baryogensis in the Early Universe. Our analysis is performed in the context of flat spacetime, but
in such a way that the main ingredients can be transferred, straightforwardly to the case of a curved spacetime of
Robertson-Walker-type. Our main technical tools are methods from constructive quantum field theory, as well as
a general method called ``projection technique''. This turns out to give convergent expansions, and a rather
elegant way of organizing the combinatorics of the various quantum field theoretic expansions in the analysis.

\end{abstract}

\pagebreak



\section{Introduction}

The Boltzmann equation is a standard tool in non-equilibrium statistical mechanics, describing the
dynamical evolution of phase space densities in a
medium with a large number of constituents such as a gas. It can be used in a wide variety
of physical contexts e.g. in order to describe transport phenomena, or the out of equilibrium dynamics
of a medium. In particular, it has been used extensively for the quantitative analysis of
abundances in early universe cosmology, such as in the context of baryogensis~(see e.g.~\cite{kolb}).

The Boltzmann equation is easy to derive heuristically, but difficult to justify from the more fundamental
viewpoint of the underlying microscopic dynamics. The standard textbook ``derivation'' (see e.g.~\cite{huang}) runs as
follows. Letting $n_k(t)$ be the average density of particles in a medium\footnote{For simplicity, we consider
 a homogeneous medium.} with momentum $k$, the infinitesimal
change $\dot n_k(t)$ arises from the collision of particles. The probability of a single collision of
two particles with incoming momenta $q_1, q_2$ and outgoing momenta
$q_1',q_2'$ that is allowed by energy-momentum conservation is given by the squared matrix element $|\M(q_1,q_2 \to q_1', q_2')|^2$. If one of the {\em initial} particles has momentum $k=q_1$, then there is a net decrease of $n_k(t)$ by the squared matrix element times the number $n_{q_1,q_2}(t)$ of colliding pairs, integrated against $dq_2$ and $dq_1' dq_2'$. If one of the {\em final} particles has momentum $k=q'_1$, then there is a similar net increase of particles. 
A key assumption is now that the numbers of colliding particles in a given volume are uncorrelated
(``molecular chaos''), i.e. that we can take $n_{q_1,q_2}(t) \sim n_{q_1}(t) n_{q_2}(t)$. With that assumption in place, 
and assuming also that the matrix element is $PT$-invariant, the
change in the particle number density at time $t$ is seen to be
\ben\label{boltzstan}
\dot n_1 = \
\Int dq_2^{} dq_1' dq_2' \,
\delta({\rm conservation \, laws}) \, |\M(q_1,q_2 \to q_1', q_2')|^2 \,
\, (n_{1'} \, n_{2'}  -  n_{1} \,  n_{2} ) \, ,
\een
where ``conservation laws'' refers to the energy-momentum conservation between incoming resp. outgoing particles.
This is the Boltzmann equation.

It is clear that a number of approximations have been made in this ``derivation''.
The first approximation was that we only need to consider collisions between two particles, not more. This assumption is
justified e.g. if the medium is very dilute. The second assumption was that of molecular chaos.
This is an assumption that does not just have to be made at an initial time, but at all times. On the other hand, the underlying microscopic laws--i.e. the Schr\" odinger equation or Newton's equations---governing the motion of the particles are deterministic in nature, i.e. determined by the initial state. Therefore, the assumption of molecular chaos is clearly an unsatisfactory one
from the theoretical viewpoint. It is something that should follow from the dynamical evolution and the initial conditions, and not something that ought to be assumed. The third crucial assumption was that the collisions occur instantaneously. One
would expect it to be justified approximately if the time scale over which there is a significant change of $n_k(t)$ is much longer than the collision time.

Thus, it is clear that the Boltzmann equation is not an exact equation, but one that will at best hold under certain approximations. In order for the dilute system approximation to work, one ought to take a limit wherein the initial densities go to zero, and wherein the time $t$ over which we observe correspondingly goes to infinity. Alternatively, one may consider the infinite time limit,
but make the assumption that the coupling constant, say $\lambda$, that determines the size of $|\M(q_1,q_2 \to q_1', q_2')|^2$
goes to zero at an appropriate rate. These are actually different scaling limits. The first one is called the
``dilute gas limit'', whereas the second is sometimes called ``$\lambda^2 t$-limit''~\cite{vhove}, because this quantity is kept constant
in the limit as $\lambda \to 0$ and $t \to \infty$. In the latter case, one expects corrections to the Boltzmann equation
due to Bose-Einstein/Fermi-Dirac statistics in the quantum case\footnote{Essentially, there will now be corrections of
the form $1 \pm n_k$ in the collision term, with $\pm$ for Bose-Einstein/Fermi-Dirac statistics.}. In order to deal with the
issue of molecular chaos, one has to analyze the validity of $n_{q_1,q_2}(t) \sim n_{q_1}(t) n_{q_2}(t)$ in these limits. This is equivalent to the question to what extent the higher order ``truncated correlation functions'' of the particle number density become small when the dilute-gas- or $\lambda^2 t$-limits are taken. Without any limits taken, the truncated correlation functions obey a complicated hierarchy (the ``BBGKY-hierarchy''\footnote{BBGKY $=$ Bogoliubov-Born-Green-Kirkwood-Yvon.}, see e.g.~\cite{bbgky}) of coupled equations, and the basic task can be viewed as showing that these equations decouple from the quantity $n_k(t)$ in the limits. Finally, one must address the issue
to what extent the finite collision time becomes irrelevant in the long-time limit.

These issues have been considered, with varying degrees of rigor and for different systems,
by a great number of researchers. Mathematical investigations in more recent times include
e.g.~\cite{hu}, who established the $\lambda^2 t$-limit in the context of a lattice fermi gas following partly ideas of
vanHove~\cite{vhove}. A different argument for
the same system was given by~\cite{salm,ey1}, who introduced an assumption that in effect truncates the BBGKY-hierarchy.
The low density limit has been considered in precise detail e.g. for quantum particles propagating in an environment with random impurities or for a quantum Lorentz gas~\cite{ben,ey2,eng,ch,chen1,chen2,salm1,sp1,sp2}.
The outcome of these investigations seems to be
generally that the Boltzmann (or the appropriate version thereof applicable to the particular model) equation is justified.

The aim of this paper is to analyze the validity of the Boltzmann equation in the context of {\em quantum field theory}. Our motivations for this investigation are the following:
\begin{enumerate}
\item
The models considered so far in a more rigorous fashion have been in the context of quantum mechanical models or lattice models. Are there any qualitatively different features for continuum quantum field theory systems with infinitely many degrees of freedom?

\item
In the $\lambda^2 t$-limit, the Boltzmann equation has been justified e.g. by~\cite{hu,salm} with the scattering matrix element in the {\em Born approximation} in the context of a fermionic lattice gas, but not for quantum field theories. More importantly, in the context of baryogensis~(see e.g.~\cite{kolb}), one frequently needs the Boltzmann equation to trace baryon number violating reactions whose net effect is invisible at the Born approximation, but
non-zero {\em loop corrections}. It is then important to understand how the incorporation of those corrections might be justified.
For example, one might ask whether one should then also incorporate any ``rescattering'' corrections at the same loop order into the Boltzmann
equation for consistency, and if so, what form might they take?

\item
For quantum field theory models on a curved space, one has additional physical effects of particle creation from the
``vacuum'' due to the expansion of spacetime (similar effects would also be present in other systems which effectively possess non-dynamical external fields). How would these be incorporated into the Boltzmann equation?

\item
Finally, one should bear in mind that the mathematical treatment of quantum field theory systems requires many different
technical tools compared to the non-relativistic systems that have been considered so far. What might be the appropriate ideas and techniques?
\end{enumerate}

In this paper, we will provide answers to questions 1), 2) and 4) in the context of a neutral Bose quantum field theory with
polynomial self-interaction in two spacetime dimensions. Our methods are specifically designed in order to make possible also
the treatment of curved spacetimes e.g. of Robertson-Walker type, but we will for simplicity only treat the case of flat
Minkowski spacetime in this paper. Question 3) will hence not be answered here, but our methods apply to this setup, which
we hope to discuss elsewhere~\cite{hl2}. Our restriction to two spacetime dimensions is mainly in order to a) be able to
apply the non-perturbative constructions available for such models that were developed mainly in the 1970's (see e.g.
the books~\cite{gj,gj1,rs} and the many references therein), and b) to avoid unessential technicalities related to UV-renormalization,
and similar difficulties with ``sharp time fields'' in dimensions $d>2$. However, if we interpret our main formulas merely in the sense of formal perturbation series,
they would be equally valid also in $d=3,4$ dimensions, but the appropriate renormalization prescription would have
to be understood. In more detail, what we do in this paper is the following:

\begin{enumerate}
\item
Our first main technical achievement is to obtain a ``pre-Boltzmann equation'', see eq.~\eqref{eq:preBeqn} in sec.~\ref{prebolz}.
This is an equation for the expected number densities $n_k(t) =
\langle N_k(t) \rangle$, where $N_k(t)$ is the time-evolved number operator. The pre-Boltzmann equation is an
exact, non-perturbative, coupled integro-differential equation for the number densities $n_k(t)$. It is of the same general
nature as the Boltzmann equation displayed above, but by contrast has an ``iterated'' collision kernel [see eq.~\eqref{B1}] on the right side.
It is also non-Markovian, in the sense that the iterated collision kernels are integrated over times in the past of $t$.
The advantage of the pre-Boltzmann equation is that it is still an exact equation (by contrast to the Boltzmann equation),
and that it is organized in such a way making it a good starting point for taking the long-time-dilute-medium limit.

The derivation of the pre-Boltzmann equation involves a
general technique called ``projection method''~\cite{rob}. In this method, one considers the time evolution operator $\e^{itH} \, . \, \e^{-itH}$ on the space of all observables,
and decomposes it into a part ``parallel'' to the space spanned by the observables $N_k$ of interest, and one ``orthogonal''.
The pre-Boltzmann equation is essentially an expansion in the ``orthogonal'' part, which is small in a suitable sense.
The essence of the projection method is recalled in sec.~\ref{sec2}, where we also give a version suitable for
time-dependent backgrounds that will be used in the sequel paper~\cite{hl2}. As we will see, one of the major advantages of
the projection method in the quantum field theoretic context is that it tames in a very elegant way the combinatorial
complexity of the various expansions.

\item We then perform a perturbative expansion of the terms in the pre-Boltzmann equation in sec.~\ref{sec5}, and we
show how each individual collision kernel may be expressed in terms of a local $S$-matrix element, which
has an expansion in terms of Feynman diagrams, see eq.~\eqref{tilMpert}. The occurrence of the this
local $S$-matrix (in the sense of Bogoliubov~\cite{b} and Epstein-Glaser~\cite{eg}), is closely tied to the fact that, in an exact description, the particle collisions will not really be complete after a finite time, but, strictly speaking, only
after an infinite amount of time. Here, we give two different but equivalent description of the collision factors. One is
in terms of retarded products, and the other is in terms of local $S$-matrix elements and corresponding position space
Feynman diagrams.

\item
The pre-Boltzmann equation is our starting point for the derivation of the actual Boltzmann equation. This is obtained
by taking the long-time-dilute-medium limit in the pre-Boltzmann equation, and this step is carried out in sec.~\ref{longtime}. Here we show that, if the densities are scaled to zero and the time is scaled to infinity at an appropriate rate, then the pre-Boltzmann equation simplifies considerably. Instead of local (in time) $S$-matrix elements, the full $S$-matrix elements now emerge, see eq.~\eqref{finalmaster}. This equation is still not the Boltzmann equation give above, because it
also involves ``rescattering terms'' that describe the effect of particles undergoing multiple collisions between the initial
time and time $t$. However, to leading order in the coupling constant $\lambda$, we obtain the Boltzmann equation as given, up to the standard kinematical factors familiar in special relativity. However, higher order corrections in $\lambda$ are also
incorporated in our equation. In particular, we find that, if one takes into account loop corrections in quantum field theory, then one should at the same time also incorporate these rescattering effects, as described systematically in eq.~\eqref{finalmaster}.

\item
In order to arrive at the main technical result of this paper, the pre-Boltzmann equation [cf. eqs.~\eqref{B1},\eqref{eq:preBeqn}] we need various
estimates of non-perturbative nature of the magnitude of the particle densities $n_k(t)$ as a function of $t$ and $k$.
These estimates require some of the machinery developed in the field of ``constructive quantum field theory'', see e.g. the
book~\cite{gj} and references therein. We mostly need the so-called ``$N$-estimates'' and ``higher order estimates'', which
in essence compare the number operators to the full Hamiltonian of the system. These estimates are derived in sec.~\ref{estimatessec}.
\end{enumerate}

Thus, the {\em main conclusions} of our paper are that 1) one can obtain a non-perturbative, exact integro-differential
equation for the number densities, which we called the pre-Boltzmann equation. 2) One can take the long-time-dilute-medium
limit of this equation, to obtain a simpler equation. This limiting equation [cf. eq.~\eqref{finalmaster}] is qualitatively similar to the Boltzmann equation, but it involves additional re-scattering terms. 3) These terms disappear to leading order in the perturbation expansion in the coupling constant (Born approximation).

We should finally indicate the levels of mathematical rigor of our arguments. The pre-Boltzmann equation is derived in a completely rigorous fashion, and it is valid non-perturbatively. The long-time-dilute-medium limit of this equation (which also
involves a thermodynamic limit) is derived under certain unproven assumptions about the existence of limits, and their
interchangeability. On the other hand, we do not drop any terms in the various expansions that appear; in particular we do not truncate the perturbation series by hand, and in this sense, our derivation is valid to arbitrary orders in perturbation theory.

\medskip
\noindent
{\bf Conventions:} Throughout the paper, we use the {\em constant convention} in our various estimates. This means that
any numerical constant is abbreviated by the same symbol, $K$, regardless of its numerical value. Thus, $K$ might
mean different constants in different inequalities. If $j,k \in \mz$, the notation $\delta(k-j)$ means the Kronecker
delta $\delta_{k,j}$, multiplied by $L$. A primed sum $\Sum_{k \in \mz}$ is an ordinary sum multiplied by $L^{-1}$.

\section{The projection method}\label{sec2}

In this section, we will outline the projection operator method, which we will use in later sections
to investigate the validity of the Boltzmann equation and its higher order corrections. This method is
well-known in the literature~\cite{rob},
and the only small new contribution in this section is only to add a generalization to the
time-dependent case. The basic framework is very general, and we will explain it,
for the sake of simplicity, in a finite dimensional situation, which has the definite advantage that
all steps are completely well-defined for elementary reasons.
In the infinite dimensional context, the calculations are formally the same,
but the convergence of the various series expression below then of course cannot be taken for granted.

Let $H \in M_n(\mc)$ be a Hamiltonian, self-adjoint on $\H = \mc^n$. Given an observable $A \in M_n(\mc)$, we define its time-evolution as usual
by
\ben
\alpha_t: M_n(\mc) \to M_n(\mc) \, \quad
\alpha_t(A) = \e^{itH} A \e^{-itH} \equiv A(t) \,\,  .
\een
The time evolution satisfies of course the group law $\alpha_{t+s} = \alpha_t \circ \alpha_s$, i.e. it is
an automorphism, and it is a homomorphism of the algebra $M_n(\mc)$ for each $t$, i.e. $\alpha_t(AB) = \alpha_t(A) \alpha_t(B)$.
In the context of the Boltzmann equation, we have a density matrix state $\rho$, i.e. an self-adjoint, positive
semi-definite operator satisfying $\tr \rho = 1$,
and a family of observables $A_j, j= 1, \dots, N$, and we want to study the time evolution of their expectation values,
\ben
a_j(t) := \tr(\rho A_j(t)) \, .
\een
The observables $A_j, j=1, \dots, N$ typically of interest are e.g. suitably defined number operators,
with $j$ corresponding to the mode number, but for the moment this is irrelevant.

One would like to derive a differential equation for the complex valued functions $a_j(t)$. This would be straightforward in principle if the set of observables (matrices) $A_j$ was a basis of $M_n(\mc)$, i.e., when $N=n^2$ and all the $A_j$ linearly independent. Indeed,
we could then simply express the linear operator $\alpha_t: M_n(\mc) \to M_n(\mc)$ as a matrix in this basis as $\alpha_t(A_j) = \sum_{k=1}^N m_{jk}(t) A_k$ for a 1-parameter group of matrices $(m_{ij}(t))$, and the desired differential equation would then simply follow by taking the expectation value of this expression and differentiating with respect to $t$. This procedure is of course not very practical nor actually different from solving the full Schr\" odinger equation, because we would need to know $m_{ij}(t)$, and this means in practice that we have to diagonalize $H$. At any rate, we will be interested in a situation where the family $A_j$ is very far from forming a basis of $M_n(\mc)$, and we consequently cannot proceed in this way. Instead, we will find an equation for the observables $a_j(t)$ that is both non-local in time and non-linear. That equation by itself is not
any simpler than the evolution equation for the operator quantities $A_j(t) = \alpha_t(A_j)$; its advantage lies rather in the fact that it is more amenable to expansion and approximation techniques, and this will eventually lead to the Boltzmann
equation.

The main tool in the derivation of this equation [see eq.~\eqref{rob0}] is the projection method.
The idea behind this method is to introduce a family of linear maps $\P_t: M_n(\mc) \to M_n(\mc)$, smooth in $t \in \mr$, with the following general properties:
\begin{enumerate}
\item We have, with $I$ the unit matrix:
$${\rm range} \, \P_t = \alpha_t[{\rm span} \{I, A_1, \dots, A_N\}] \subset M_n(\mc).$$
\item We have
$$\P_t \circ \alpha_{t-s} \circ \P_s = \alpha_{t-s} \circ \P_s$$ for any $t \ge s$.
\end{enumerate}
The family of maps $\P_t, t \in \mr$ is referred to as a family of ``projections'' onto the space of time-$t$ observables, even though they are in fact just idempotents, $\P_t^2 = \P_t$, and not
projections according the standard terminology in functional analysis (we are not giving the space $M_n(\mc)$ any hermitian structure, so there is no sense in which the $\P_t$ are self-adjoint). We neither require the maps $\P_t: M_n(\mc) \to M_n(\mc)$ to be algebra homomorphisms.  The projections will serve
to break up the time evolution of an observable into a part ``parallel'' to $\P_t$, and a complementary
part ``parallel'' to the complementary projection $\Q_t := id - \P_t$, where $id: M_n(\mc) \to M_n(\mc)$ is the identity.
Later, $\P_t$ will be chosen in such a way that
the latter part becomes small in a suitable sense, and this will then be treated as a perturbation.

To start, and to simplify our notation, we note that the Heisenberg evolution equation for a matrix
$X \in M_n(\mc)$ that comes from $H$ can be written
simply as
\ben
\frac{d}{dt} \alpha_t(X) = i \, \delta \alpha_t(X) \, , \quad \delta(X) = [H,X] = HX-XH \, .
\een
We
decompose
\begin{equation}\label{eq:ProjOp_1}
\frac{d}{dt} \alpha_{t} = i \alpha_t \circ \delta = i \P_t \circ \alpha_t \circ \delta + i \Q_t \circ \alpha_t \circ \delta
= i \alpha_t \circ \tilde \P_t \circ \delta + i \alpha_t \circ \tilde \Q_t \circ \delta \, ,
\end{equation}
where we have found it convenient to introduce the ``Schr\" odinger picture''
operators $\tilde \P_t = \alpha_{-t} \circ \P_t \circ \alpha_t$.
The goal is now to replace $\alpha_t \circ \tilde \Q_t \circ \delta$
with an expression that involves only terms with the map $\tilde \P_t$. To this end one
notes  that the following differential equation
\begin{equation}\label{eq:ProjOp_2}
\begin{split}
\frac{d}{dt} \left[ \alpha_{t} \circ \tilde \Q_t \right] &=
i\alpha_{t} \circ \delta \circ \tilde \Q_{t}
+ \alpha_{t} \circ \frac{d}{dt} \tilde \Q_t  = \\
&= i\alpha_{t} \circ \tilde \P_t \circ \delta \circ \tilde \Q_t +
  i\alpha_{t} \circ \tilde \Q_{t} \circ \delta \circ \tilde \Q_t +
  \alpha_{t} \circ \frac{d}{dt} \tilde \Q_t \, .
\end{split}
\end{equation}
This is an inhomogeneous linear differential equation for the
operator $\alpha_{t} \circ \tilde \Q_t: M_n(\mc) \to M_n(\mc)$, which we can therefore integrate
straightforwardly, with the result
\begin{equation}\label{eq:ProjOp_3}
\alpha_{t} \circ \tilde \Q_t =
i \tilde \Q_0 \circ \tilde \E_{0,t} \circ \delta +
\Int_{0}^t \alpha_{s} \circ
\left[ i\tilde \P_{s} \circ \delta \circ \tilde \Q_s +
\frac{d}{ds}\tilde \Q_s \right] \circ \tilde \E_{s,t} \, ds \, .
\end{equation}
Here $\tilde \E_{s,t}: M_n(\mc) \to M_n(\mc)$ is the cocycle (meaning that $\tilde \E_{t_1, t_2} \circ \tilde \E_{t_2,t_3} = \tilde \E_{t_1,t_3}$)
defined to be the solution to the homogeneous differential equation
\ben
\frac{d}{dt} \tilde \E_{s,t}(X) = i \, \tilde \E_{s,t} \circ \delta \circ \tilde \Q_t(X) \, , \qquad \tilde \E_{s,s}(X) = X \, ,
\een
for all $X \in M_n(\mc)$. The solution can be written as $\tilde \E_{s,t} = \alpha_{-s} \circ \E_{s,t} \circ \alpha_t$, where $\E_{s,t}$ is the ``Heisenberg-picture cocycle'' given the summation formula
\ben\label{eseries}
\E_{s,t}(X) = \sum_{k=0}^\infty (-i)^k \Int_{t>\sigma_k>...>\sigma_1>s} \delta \circ \P_{\sigma_1} \circ \cdots
\circ \delta \circ \P_{\sigma_k}(X) \, d^k\sigma \, .
\een
Note that this sum trivially converges as we can estimate it by
\ben
\| \E_{s,t}(X) \| \le
 \|X\| \sum_{k=0}^\infty
 \frac{|t-s|^k}{k!} \| \delta \|^k (\sup_\sigma \| \P_\sigma \|)^k \le \e^{K|t-s|} \|X\| \, ,
\een
since the volume of the set $\{t>\sigma_k>...>\sigma_1>s\}$ is given by the first term under the summation sign.
Switching from the ``tilde'' projectors back to the original ones, equation~\eqref{eq:ProjOp_1} now takes the form
\begin{equation}\label{eq:eqn_motion_decomposed}
\begin{split}
\frac{d}{dt} \alpha_{t}(X) & = i\P_t \circ \alpha_t \circ \delta(X)
                   + i\Q_0 \circ \E _{0,t} \circ \alpha_t \circ \delta(X) +  \\
                 & \qquad \qquad + i \Int_{0}^t \left[
          \P_s \circ \delta \circ \Q_{s} -
          \alpha_s \circ \frac{d}{ds} (\alpha_{-s} \circ \P_s \circ \alpha_{s}) \circ
          \alpha_{-s} \right] \circ \E_{s,t} \circ \alpha_t \circ \delta(X) \, .
\end{split}
\end{equation}
This equation is the starting point for our analysis. As it is, it is--if anything--more
 complicated than the Heisenberg equation of motion that we started with, in particular
 it is an integro-differential of motion, rather than a differential
 equation. But it will later be seen that it is useful to
 study approximations.

So far, we have not yet made a specific choice for our projectors, $\P_t$. A particularly useful choice
is available when the observables $A_j, j=1, \dots, N$ mutually commute, i.e., $[A_i, A_j] = 0$ for
all $i,j$, and when all of them are hermitian, $A_j = A_j^*$ for all $j$. We will assume this from now.
We wish to define, for each $t \in \mr$ a reference state $w_t: M_n(\mc) \to \mc$ which reproduces the expectation values of
the observables $A_j(t)$ in the given state $\rho$. Thus, $w_t$ should be a linear functional which is
normalized and positive in the sense that $w_t(I) = 1, w_t(X^*X) \ge 0$ for all $X$, and for which
\ben\label{wt00}
w_t(A_j(t)) = a_j(t) \, ,
\een
for all $j$. It is easy to see that there will be, in general many, solutions to this equation. Indeed, dropping the
reference to ``$t$'' for simplicity, let
${\mathcal A} = {\rm alg}\{A_j, j=1, \dots, N\}$ be the abelian $*$-algebra generated by the observables.
By making a joint spectral decomposition $A_j = \sum_{\alpha = 1}^n f_j(\alpha) p(\alpha)$, with
$p(\alpha)$ rank-1 projections, we can identify $\mathcal A$ with a subalgebra of the abelian $*$-algebra of functions
${\rm Fun}(\{1, \dots, n\} \to \mc)$, and we can identify $\tr(\rho \, . \,)$ with a positive linear functional
on this algebra. By standard theorems, there is then a non-negative function
$m: \{1, \dots, n\} \to \mr$ of total weight one such that $a_j = \sum_{\alpha=1}^n f_j(\alpha) \, m(\alpha)$.
We define our state (reintroducing the dependence on $t$) e.g. by
$w_t(X)= \tr[X \sum_{\alpha = 1}^n p_t(\alpha) m_t(\alpha)]$. This is usually not the only solution to eq.~\eqref{wt00}.
It is convenient to take a solution which maximizes the ``entropy'', i.e. to take
\ben
\label{wtdef}
w_t(X) = \tr(\rho_t X) \, , \qquad \tr \rho_t = 1 \, , \quad \rho_t \ge 0
\een
in such a way that the functional
\ben\label{entropy}
S[\rho_t] = -\tr(\rho_t \log \rho_t)
\een
is maximized. A standard argument involving Lagrange multipliers shows that, in the generic case\footnote{\label{f1}
Generic means here that the maximizer is strictly inside the convex set of all positive linear functionals
on $M_n(\mc)$.}, this
maximizer must be of the form
\ben
\rho_t = \frac{1}{Z(t)} \exp \left( - \sum_{j=1}^N \mu_j(t) A_j(t) \right) \, .
\een
Let us assume that the non-negative functions $\mu_j: \mr \to \mr$ have been
chosen in this way, i.e. that we are in the ``generic case'' for all $t$. In our applications
below, the nature of the observables $A_j$ implies that the functionals $w_t$ always exist, and we will
simply assume the same here.
We can then use the reference states $w_t$ in order to construct the projection operator in question. Let us
define the ``correlation $N\times N$ matrix'' as usual by
\ben
c_{ij}(t) = w_t[(A_i(t) - a_i(t) I)(A_j(t) - a_j(t) I)]  \, .
\een
This matrix is positive semi-definite, and generically~\footnote{The same remark as in footnote~\ref{f1} applies here.}
invertible. We denote the inverse as $c^{ij}(t)$. We now
define our projection operator $\P_t$ as
\ben\label{projdef}
\P_t(X) = w_t(X) I + \sum_{i,j=1}^N c^{ij}(t)  \,\, w_t[X(A_j(t)-a_j(t)I)] (A_i(t) - a_i(t) I)\, .
\een
The following lemma is an immediate consequence of this definition:
\begin{lemma}
For any $X \in M_n(\mc)$ any $t \in \mr$, we have
\ben
\alpha_t \circ \frac{d}{dt} (\alpha_{-t} \circ \P_t \circ \alpha_t(X)) \in
{\rm span} \left\{ A_j(t) - a_j(t)I \mid j = 1, \dots, N \right\} \, .
\een
For any
$Y \in {\rm span} \, \{I, A_1(t), \dots, A_N(t)\}$, we have $\P_t \circ \delta(Y) = 0$.
We also have $w_t(X) = \tr (\rho \P_t(X))$ for any $X \in M_n(\mc)$.
\end{lemma}
\medskip\noindent
{\em Proof:} To prove the first statement, we introduce $\tilde \P_t = \alpha_{-t} \circ \P_t \circ \alpha_t$ as above,
and we also introduce $\tilde w_t = w_t \circ \alpha_t$. The first statement is then seen to be equivalent
to the statement that $\frac{d}{dt} \tilde \P_t(X) \in {\rm span} \{A_j - a_j(t) I \mid j=1,\dots,N \}$. Now we have,
using the ``summation convention'':
\ben
\tilde \P_t(X) = \tilde w_t(X) + c^{ij}(t) \, \tilde w_t[X (A_i - a_i(t)I)] (A_j - a_j(t)I) \, .
\een
When taking the $t$-derivative of this expression, we note the identities (dropping the reference to $t$):
\ben\label{trick}
\frac{\partial a_i}{\partial \mu_j} = c_{ij} \, , \quad \frac{\partial \mu_j}{\partial a_i} = c^{ij} \, ,
\een
from which it immediately follows that
\ben
\frac{d}{dt} \tilde w_t(X) = \tilde w_t[X (A_i - a_i(t)I)] \, \frac{\partial \mu_i}{\partial a_j}(t) \, \dot a_j(t) =
c^{ij}(t) \, \tilde w_t[X (A_i - a_i(t)I)] \, \dot a_j(t) \, .
\een
This term cancels precisely the derivative of the second term in $\frac{d}{dt} \tilde \P_t$ when
the derivative hits $a_j(t)$ in that term. The remaining terms are given by a linear combination of
$A_j - a_j(t)I$, as claimed.

In order to prove the second statement, we just follow the definitions and use the cyclic property of the trace
as well as the fact that each $A_j(t)$ commutes with the density matrix $Z(t)^{-1} \exp-\sum \mu_j(t) A_j(t)$.
The last statement is again a straightforward consequence of the definitions. \qed

\medskip
\noindent

We are now almost ready to derive the desired integro-differential equation for the $a_j(t)$.
To obtain a particularly simple form, we shall make the {\bf initial state assumption}
that at $t=0$ we have
\ben\label{initial}
w_0(X) = \tr(\rho X) \quad \text{for any $X \in M_n(\mc)$;}
\een
in other words that $\rho = Z^{-1} \exp (-\sum \mu_j A_j) = \rho_0$ for some $\mu_j = \mu_j(0) \in \mr$.
The physical meaning of this hypothesis will be explained below. With this assumption in place,
we proceed as follows: We take the expectation value in our density matrix state $\tr(\rho \, . \, )$ for
$X = A_j$ in eq.~\eqref{eq:eqn_motion_decomposed}. Then the first term on the r.h.s. is seen to disappear
using the last two statements of the previous lemma. The second term on the r.h.s. is seen to disappear
when acted upon by $\tr(\rho \, . \,)$ using the last statement in the lemma, and the assumption
eq.~\eqref{initial} on the initial condition. For the first term under the integral, we use
the preceding lemma repeatedly to write
\ben
\begin{split}
\tr[\rho \, \P_s \circ \delta \circ \Q_s \circ \E_{s,t} \circ \alpha_t \circ \delta(A_j)]
&= w_s[\delta \circ \Q_s \circ \E_{s,t} \circ \delta (A_j(t))]\\
&= w_s[\delta \circ \E_{s,t} \circ \delta (A_j(t))] \, .
\end{split}
\een
Finally, the last term under the integral disappears when acted upon by $\tr(\rho \, . \,)$, because
it is in the span of $A_j(s) - a_j(s)I$, again by the preceding lemma, and this is annihilated by
$\tr(\rho \, . \,)$. Thus we arrive at the following theorem:

\begin{thm}\label{robertson}
Let $A_j, j=1, \dots, N$ be a set of hermitian, mutually commuting complex $n \times n$ matrices, let $\rho$ be a density matrix
(self adjoint positive definite matrix of unit trace) of the general form $Z^{-1} \exp (-\sum \mu_j A_j)$. Let
$a_j(t) = \tr(\rho A_j(t))$, where $A_j(t) = \e^{itH} A_j \e^{-itH}$ is the time evolved observable with respect to
a self-adjoint hamiltonian $H$, and
let $\E_{t,s}: M_n(\mc) \to M_n(\mc)$ be the cocycles defined as above in eq.~\eqref{eseries}.
Then the equation
\begin{equation}\label{rob0}
\frac{d}{dt} a_j(t) = -\Int_{0}^t
          w_s \left[ \delta  \circ \E_{s,t} \circ \delta(A_j(t)) \right] \, ds \,
\end{equation}
holds, where $\delta(X) = HX-XH$. For an arbitrary density matrix state $\rho$, i.e. if we do not make
the initial state assumption~\eqref{initial}, the equation takes the form
\ben\label{rob1}
\frac{d}{dt} a_j(t) = (\rho-w_0) \left[\E _{0,t}\circ\delta (A_j(t)) \right] -\Int_{0}^t
          w_s \left[ \delta  \circ \E_{s,t} \circ \delta(A_j(t)) \right] \, ds \,.
\een
Here, we are using the shorthand $(\rho-w_0)(X) = \tr(\rho X)-w_0(X)$ for any $X$.
\end{thm}

The equation in the above theorem is known in the literature as ``Robertson equation''. The initial state assumption~\eqref{initial} on the density matrix state $\tr(\rho \, . \,)$ is made mainly for convenience. Its physical
interpretation is that we have ``maximal ignorance'' about the initial state, because saying that
$\tr(\rho \, . \,) = w_0$ means that the initial state is the one with maximal entropy~\eqref{entropy},
among all the states with prescribed initial values $a_j(0)$.

The initial state condition can be dropped at the expense of another term in the Robertson
equation which encodes the corresponding initial state, see eq.~\eqref{rob1}. This term describes the way in which the
influence of the chosen initial state persists to later times. In the model studied below,
$A_j$ will be the number operators at the initial time, with $j$
a mode number. In that case, the initial condition can be viewed as saying that the initial state is quasifree, or in a sense,
as uncorrelated as possible. This is a physically reasonable assumption, since what one wants to study is not the effect of strong correlations
persisting from the initial state to later times, but the process of approach to equilibrium.

Our derivation has the
advantage that it can be transferred, relatively straightforwardly, to the case when the dynamics is given by
a time-dependent hamiltonian $H(t)$, (smooth in $t$, say). This situation will be of interest to us in~\cite{hl2}, because
we want to apply the formalism to field theory in curved (time dependent) backgrounds.
Let us explain briefly the changes that have to be made to the statement and proof of thm.~1 in that situation.
For a time-dependent hamiltonian, the derivation $\delta_t(X) = H(t) X - X H(t)$ depends on time, and the time evolution
operator is now a cocycle $\alpha_{t,s}$ satisfying $\frac{d}{dt} \alpha_{t,s} = i\delta_t \circ \alpha_{t,s}$,
and a time evolved observable is $A(t) = \alpha_{t,0}(A)$. The
functional equation for the projection maps is replaced by $\alpha_{t,s} \circ \P_s = \P_t \circ \alpha_{t,s} \circ \P_s$. The formula
for $w_t$ is the same as before and the cocycle is given now by
\ben\label{eseries1}
\E_{s,t}(X) = \sum_{k=0}^\infty (-i)^k \Int_{t>\sigma_k>...>\sigma_1>s} \delta_{\sigma_1} \circ \P_{\sigma_1} \circ \cdots
\circ \delta_{\sigma_k} \circ \P_{\sigma_k}(X) \, d^k\sigma \, .
\een
With those changes understood, the Robertson equation
remains the same as above up to the obvious changes:
\begin{equation}\label{eq:eqn_motion_decomposed1}
\frac{d}{dt} a_j(t) = -\Int_{0}^t
          w_s \left[ \delta_s  \circ \E_{s,t} \circ \delta_t (A_j(t)) \right] \, ds \, \, .
\end{equation}

\section{The $\phi^{p}$-model in two dimensions}\label{sec3}

\subsection{Basic features of the model}

In this paper, we will study the Boltzmann equation in the context of a particular quantum field theory model,
describing a single hermitian Bose scalar field in two spacetime dimensions interacting only with itself, with a polynomial self interaction.
The mathematical properties of these models have been well studied in the literature of the 1970's (see e.g. the book~\cite{gj,gj1}).
In particular, it has been demonstrated that the model exists, in a non-perturbative sense, and that it satisfies
the usual properties expected from a quantum field theory on general grounds. The purpose of this section is to
review some of the basic constructions and theorems for these models (in the ``operator-'' rather than ``path-integral''
approach), and to introduce the quantities of main interest in this paper, namely the expected number densities $n_k(t)$.
We then proceed to derive various bounds on these quantities that are needed in the following sections. Here
we rely to a large extent on established techniques in the study of this model, namely the ``higher order estimates'' (which we call``Rosen inequalities''), the ``$N$-estimates'',
as well as the position space representation.

We will begin by defining this model on a spatially compact spacetime, i.e. $\mr \times S^1$, with metric
\ben
ds^2 = -dt^2 + dx^2 \, .
\een
Here $x$ is a $2\pi L$-periodic coordinate parameterizing $S^1$, so
that the circumference of the spatial $S^1$ is $2\pi L$. The model may be characterized by writing its Hamiltonian,
given by
\ben\label{Hdef}
\begin{split}
H &= \frac{1}{2} \Int_0^{2\pi L} :\left(  \Pi(x)^2 + \left[ \frac{d\phi(x)}{d x} \right]^2 + m^2 \, \phi(x)^2 + \lambda \sum_{n=0}^p b_n \, \phi(x)^n \right) : \, dx \\
& \vspace{2mm}\\
&\equiv H_0 + \lambda V \, .
\end{split}
\een
Here, $\phi,\Pi$ are fields satisfying (in a suitable weak sense) the canonical commutation relations
$[\Pi(x),\phi(y)] = i \delta(x-y)$, and the double dots denote normal ordering, defined more precisely
below. We require that $m>0$ and that $\lambda \ge 0$, and we assume that the polynomial $P(\xi) = \sum b_i \xi^i$ is
non-negative\footnote{Obviously, $p$ then has to be an even number.}, i.e. $P(\xi) \ge 0$ for all $\xi \in \mr$ and even, but otherwise arbitrary.
The coupling constants of the model are hence $m,\lambda$ and the coefficients $b_i$. Strictly speaking, $\lambda$
is redundant and could be absorbed into the coefficients, but we keep it explicitly because it is simpler to have one,
instead of many, expansion parameters below.  The fact that $m>0$ implies~\cite{gj} that, in the IR-limit $L \to \infty$,
the suitably shifted Hamiltonian $H$ has a
state of lowest energy, followed by has a mass gap, corresponding to physical particles in the sense
of scattering theory. However, we note that the value of this mass gap is {\em not} $m$, and hence this parameter must therefore {\em not} be confused with the physical mass.

For the mathematical construction of the model, it is essential that the Hamiltonian $H$ can be rigorously defined as a self-adjoint operator on a Hilbert-space $\H$, and for stability it is essential that $H$ is bounded as an operator from below by a constant $-O(\lambda)L \cdot I$ when $\lambda  \ge 0$. This is a highly non-obvious fact, because, although we have that $\lambda P(\xi) \ge 0$ for any $\xi \in \mr$, the potential $V$ is {\em not} positive definite as an operator but in fact even unbounded from below. The latter is an unavoidable consequence of the normal ordering prescription without which the expression for $V$ would be ill-defined.
The point is however that the sum $H = H_0 + \lambda V$ is bounded from below by a constant
$-O(\lambda)L$ (times the identity operator $I$)~\cite{gj}, i.e. we have that $H \ge -O(\lambda)L \cdot I$ in the sense of operators.
This is essentially because one can show that for states $\Psi \in \H$ for which $(\Psi, V \Psi)$ becomes very negative,
the contribution $(\Psi, H_0 \Psi)$ becomes very positive, and in effect overcompensates the negative contribution from the potential $V$.
We are going to redefine $H$ by the constant $O(\lambda) L \cdot I$. This does not affect the definition of the time evolution $\alpha_t(A)
= \e^{itH} A \e^{-itH}$ of
  an observable $A$, but it has the advantage of turning $H$ into a non-negative operator.

Because $H$ is essentially self adjoint, we can define in a mathematically unambiguous way the time evolution operators
$\e^{itH}$, and we can then also define the spacetime field operators $\phi(t,x) = \e^{itH} \phi(x) \e^{-itH}$ as operator valued distributions on $\H$, i.e., if $f \in C^\infty_0(\mr \times S^1)$, then $\phi(f) = \int \phi(t,x) f(t,x) \, dtdx$
is essentially self-adjoint e.g. on the domain $\cap_j \D(H^j_0) \subset \H$.
The creation and annihilation operators of the model can be defined for $k \in \mz$ as
\ben\label{akt}
a_k(t) = \Int_0^{2\pi L} [\phi(t, x) \partial_t u_k(t,x) - u_k(t, x) \partial_t \phi(t,x)] \, dx \, ,
\een
where the ``positive frequency mode functions'' are defined as
\ben\label{ukt}
u_k(t,x) = \frac{1}{(2\pi)^{\half} (2\omega_k)^{\half}} \e^{i\omega_k t - ikx/L} \, , \quad \omega_k = \sqrt{k^2/L^2 + m^2} \, .
\een
The $a_k(t)$ are well-defined as quadratic forms on $\H$ with domain e.g. $\D(H) \times \D(H)$, and satisfy the standard
algebra
\ben
[a_k(t), a_p(t)^*] = \delta(p-k) \, , \quad [a_k(t), a_p(t)] = 0 = [a_k(t)^*, a_p(t)^*] \, ,
\een
where $\delta$ is the {\em scaled} Kronecker-delta, defined by
\ben
\delta(k-p) = \begin{cases}
L & \text{if $k=p$,}\\
0 & \text{otherwise.} \end{cases}
\een
In the absence of self-interaction (i.e. when $\lambda = 0$), the annihilation and creation operators $a_k^\#(t)$ are
independent of $t$, but otherwise they are time-dependent. In terms of the creation and annihilation operators, the (unshifted) Hamiltonian is
\ben
H = \Sum_{k \in \mz} \omega_k^{} a_k^* a_k^{} +  \pi\lambda \sum_{n=0}^p  \frac{b_n}{(2\pi)^{\frac{n}{2}}} \Sum_{X \cup Y = \{1, \dots, n\}} \delta \left( k_X-k_Y \right) \prod_{j \in Y} \frac{a_{k_j}^*}{(2\omega_{k_j})^{\frac{1}{2}}} \prod_{i \in X} \frac{a_{k_i}^{}}{(2\omega_{k_i})^{\frac{1}{2}}} \, .\non
\een
Here we are using the shorthand $k_X = \sum_{i \in X} k_i$ etc., and a primed sum over $k \in \mz$ is defined to be the {\em rescaled} sum $\Sum_{k \in \mz} := \frac{1}{L} \sum_{k\in \mz}$, with multiple sums having correspondingly more inverse powers of $L$. In the expression for $H$, we can change $a_k^\#$ for $a_k^\#(t)$, i.e. we can take the creation and annihilation operators at any time, because $H$ is a constant of motion. But this is not so for the free hamiltonian $H_0$ and the interaction term $V$. The normal ordering prescription referred to by double dots in eq.~\eqref{Hdef} is equivalent to the fact that all creation operators stand to the left of all
annihilation operators in the above expression.

Of course in practice, the above expression for $H$ in terms of creation and annihilation operators is the starting point of the analysis. One defines $\H$ to be the standard Bosonic Fock space,
\ben
\H = \mc \oplus \bigoplus_{n=1}^\infty {\mathbb P}_n [\underbrace{\ell^2(\mz) \otimes \cdots \otimes \ell^2(\mz)}_{n}]
\een
where ${\mathbb P}_n$ projects onto the subspace of totally symmetric rank $n$ tensors over $\ell^2(\mz)$, the 1-particle Hilbert
space of square summable sequences, and the summand $\mc$ stands for the ``vacuum'' vector $\Omega_0$ (note that
this is very different from the true ground state of $H$!). The
scalar product is that inherited from\footnote{The inner product on this $\ell^2$ is
defined for convenience with the scaled sum $\Sum_{k \in \mz}$.} $\ell^2(\mz)$, and the creation operators are defined on an $n$-particle state
$\Psi_n$ as $a_k^* \Psi_n = {\mathbb P}_{n+1} (\delta(k- \, . \,) \otimes \Psi_n)$, where $\delta(k - \, . \,) \in \ell^2(\mz)$ is the sequence consisting of $0$'s and precisely one $L$ in the $k$-th place.

\medskip

In this work, the observables of main interest are the number density operators
\ben
N_k = \frac{1}{L} a_k^* a_k^{} \, ,
\een
and their time evolved expectation values. In the analysis of these quantities, we will frequently need to compare the number density operators
or the free Hamiltonian $H_0$ to the interacting Hamiltonian $H$. Such inequalities (``higher order estimates'') have
been given e.g. by Rosen~\cite{rosen}:

\begin{thm} (``Rosen's inequality'')  For each natural number $j$ and each $\epsilon>0$,
there is a natural number $i$ depending on $\epsilon,j$ and the degree $p$ of the interaction polynomial
such that
\ben
N^{j+\epsilon-3} H_0^{3-\epsilon} \le \c (H + O(\lambda) L \cdot I)^i
\een
where $N = \sum_{k \in \mz} N_k$ is the total number operator, and where
 the constant $K$ does not depend upon $\lambda$ or $j$. If $\epsilon>2$,
 we may take $i=j$.
\end{thm}

 We will use this inequality in many places below, and we will, for simplicity
 absorb the additive constant $O(\lambda) L \cdot I$ into the Hamiltonian $H$. A simple, but important observation is
  that same inequality applies to the time evolved free Hamiltonian and time evolved number operator, because
  $H$ is a constant of motion. In this way, the higher order estimates will allow us to
  transfer information on the number densities at the initial time to later times.

For some of our arguments below, it is also convenient to introduce yet another Hilbert space representation--essentially the $\QQ$-space representation---on the space $L^2(\QQ, d\nu)$, see~\cite{gj} for details. Here $(\QQ,d\nu)$ is the
measure space given (formally) by infinitely many Cartesian copies of $\mc \cong \mr^2$ and Gaussian probability measure
of total weight one,
\ben
d\nu  = \prod_{j \ge 0} \pi^{-1} \e^{-|q_j|^2} \, dq_j d \bar q_j\, .
\een
On this space we have
canonical (complex) multiplication and differentiation operators
\ben
P_k \Psi = -i\e^{|q_k|^2/2} \frac{\partial}{\partial q_k} (\e^{-|q_k|^2/2} \Psi) \, , \quad
Q_k \Psi = q_k \Psi \, ,
\een
where $q_k \in \mc, k \ge 0$. They satisfy the standard canonical commutation relations
$[Q_k, P_j] = i\delta_{k,j}$. The Hilbert space $\H$ is related to the space of square integrable functions on $\QQ$
by an isometry $W: \H \to L^2(\QQ, d\nu)$. This isometry relates the operators $P_k, Q_k$ to the
creation and annihilation operators introduced above  by
\ben
W^* Q_k W = \frac{1}{\sqrt{2\pi L}} (a_k^* + a_{-k}^{}) \, , \quad
W^* P_k W = \frac{i}{\sqrt{2\pi L}} (a_k^* - a_{-k}^{}) \, .
\een
The ``vacuum'' vector in $\H$ is mapped to $W \Omega_0 = 1$, the identity function in $L^2$. This together
with the relations just given uniquely determines $W$; its action on states with higher particle number
gives products of Hermite-polynomials in the variables $q_k$. The advantage of the $\QQ$-space representation
is that the interaction $V$ turns into a multiplication operator. In fact, the  Hamiltonian reads in
this representation
\ben
W^*H W = \frac{1}{2} \sum_{k \in \mz} \omega_k (|P_k|^2 + |Q_k|^2 -1) + \lambda V(Q_0,Q_1,\dots) \, .
\een
Here, $V$ is defined as $V = \lim_{\Lambda \to \infty} V^\Lambda$, where we have set $Q_{-k} = Q_k^* = \bar q_k$
for $k \ge 0$ (and similarly for $P_{-k}$), and where
\ben
V^\Lambda(q_0, q_1, \dots) =  \pi \sum_{n=0}^p \,\, b_n^\Lambda \,\, \sum_{|k_1|, \dots, |k_n| \le \Lambda}  \delta(\sum_j k_j)
\, \frac{q_{k_1} \cdots q_{k_n}}{(2\pi L\omega_{k_1})^{\half} \cdots (2\pi L\omega_{k_n})^{\half}} \, .
\een
We have also introduced $b_j^\Lambda = \sum_{i} (-1)^i \frac{(j+2i)!}{j!i!} c_\Lambda^i b_{j+2i}$,
with $c_\Lambda = (2\pi L)^{-1} \sum_{|k| \le \Lambda} (2\omega_k)^{-1} \sim \log \Lambda$. It can be shown that
$V(q_0, q_1, \dots)$ is in $L^p$ for any $p< \infty$, see e.g.~Prop.~2.1.2 of~\cite{gj},
where we mean the usual the H\"older space $L^p(\QQ,d\nu)$. It is the Banach space with the norm
$\|\Psi\|_p = (\int |\Psi|^p \, d\nu)^{1/p}$. From now on we will drop the reference to $W$.

One can use the $\QQ$-space representation e.g. to give a relatively simple proof of the well-known fact that $\e^{-\beta H}$ is a trace class operator.
Since we will occasionally appeal to this result, we state it as a
\begin{lemma}\label{trace}
We have $\tr (\e^{-\beta H}) \le \e^{\c \beta^{-1}} < \infty$ for any $\beta>0$.
The same applies to $\e^{-\beta H - \sum \mu_k N_k}$, where $\mu_k \ge 0$.
\end{lemma}
\medskip
\noindent
{\em Remark:} A careful look at the estimates in the proof shows that the constant $\c$ is
of order $L$, so the free energy goes as $\beta^{-1} L$, as one expects.

\medskip
\noindent
{\em Proof:} We give the proof without the ``chemical potentials'' $\mu_k$ for simplicity.
The general case is the same because the $\mu_k$ can be absorbed into the $\omega_k$ in the free
Hamiltonian, and this only makes things better.

The Golden-Thompson inequality states that $\tr (\e^{A+B}) \le \tr (\e^{A} \e^B)$ for any
hermitian matrices $A,B$. The inequality can be proved e.g. using standard properties of the trace
and the Lie-Trotter product formula
\ben
\lim_{n \to \infty} (\e^{A/n} \e^{B/n})^n = \e^{A+B} \, .
\een
It is possible to apply this kind of reasoning also in the infinite dimensional context
to $A= H_0, B = \lambda V$. Indeed, an appropriate version of the Lie-Trotter formula~(see e.g.~\cite{rs}) then holds,
because $H$ is essentially self-adjoint on the domain ${\mathscr D}(H_0) \cap {\mathscr D}(V)$, see e.g.~thm.~3.2.1
of~\cite{gj}. From the Golden-Thompson inequality, we then get, using also the operator inequality
$\tr (XY) \le \|X\| \tr |Y|$, and denoting by $\| \, . \, \|_{p,q}$ the norm of an operator $L^q \to L^p$:
\ben
\begin{split}
\tr \, \e^{-\beta H} &\le \tr (\e^{-\beta H_0} \e^{-\beta \lambda V}) \le \| \e^{-\beta \lambda V} \e^{-(\beta/2)H_0} \|
\cdot \tr (\e^{-(\beta/2) H_0}) \\
&\vspace{0.2cm}\\
&\le \e^{\c \beta^{-1}} \| \e^{-\lambda \beta V} \|_{(2+2\c \beta)/(\c \beta)} \| \e^{-(\beta/2)H_0}\|_{2+2\c \beta,2} \non\\
&\vspace{0.2cm}\\
&\le \e^{\c \beta^{-1}} \|\e^{-[2\lambda(1+\c \beta)/K]V} \|^{\c \beta/(2+2 \c \beta)}_1  \le \e^{\c\beta^{-1}} \, .
\end{split}
\een
Here we have used that $\e^{-tH_0}$ is a contraction between $L^2 \to L^{2+2\c t}$ for some constant $\c>0$ when
$m>0$ (see thm.~2.2.5 on p.~36 of~\cite{gj}), which can in principle be seen from the well-known
explicit formula in $\QQ$-space, given by
\ben
(\e^{-tH_0} \Psi)(q) = \int_{\mathcal Q} \prod_{j \in \mz} (1-\e^{-t\omega_j})^{-\half} \, \exp \left( -\frac{|q_j'-\e^{-\omega_j t} q_j|^2}{2(1-\e^{-2\omega_j t})} + \frac{1}{2} |q_j'|^2 \right) \, \Psi(q') \, d\nu(q') \, ,
\een
where $q_{-k} = \bar q_k$ for $k \ge 0$. We are also using that $\e^{-\beta V}$ is a multiplication operator in $\QQ$-space
 whose $L^p$-norms are all finite for $1<p<\infty$ (see Thm.~2.1.4 on p.~30 of \cite{gj}), and we have used the standard inequality
 $\tr(\e^{-(\beta/2) H_0} ) \le   \e^{\c \beta^{-1}}$.
 \qed

\subsection{Estimates on the number densities}\label{estimatessec}

We now discuss properties of the expected number densities in suitable states. For $k \in \mz$, the time-evolved
number densities are given by
\ben\label{Nkt}
N_k(t) = \e^{itH} N_k \e^{-itH} = \frac{1}{L} a_k(t)^* a_k(t) \, ,
\een
where the second equality follows by a simple and straightforward calculation
noting e.g. that $\frac{d}{dt} a_k(t) = i[H,a_k(t)] - i\omega_k a_k$. The expectation values of these observables $(k \in \mz)$ in a density matrix state $\rho$ are denoted
\ben
n_k(t) = \tr [ \rho N_k(t) ] \, , 
\een
where we must require at least that $\rho \in {\mathscr I}_1(\H)$, the space of all trace-class operators on $\H$.
These quantities depend on $\lambda$ because the Hamiltonian $H(\lambda) = H_0 + \lambda V$ does. We will give estimates
concerning the magnitude of $n_k(t)$ as a function of the mode number, time, and the coupling constant $\lambda$.
These will be used later when we investigate the
``pre-Boltzmann equation''. Our main estimates are contained in the following theorem:
\begin{thm}\label{nestim}
Assume that $\rho H^j \in {\mathscr I}_1(\H)$ up to sufficiently large $j$,
and let $n_k(t,\lambda) = \tr (N_k(t) \rho)$ as above, with $\rho$ independent of $\lambda$.
Then for $\half>\epsilon>0$ we have the bounds
\ben
|n_k(t,\lambda) - n_k(s,\lambda)| \le \c \lambda |t-s| \omega^{-1-\epsilon}_k \,
\een
and
\ben
|n_k(t, \lambda)| \le \c \omega_{k}^{-4+\epsilon} \, ,
\een
with a constant depending on $\epsilon, p$ and $L$.
\end{thm}
{\em Remark:} The proof shows that the constants are of the order of $\tr (\rho H^j)$ where $j$ is
some positive number depending on $\epsilon$. This quantity is not uniformly bounded in $L$ (in
fact typically it is $\propto L^j$), hence the estimates are not guaranteed to be
preserved in the thermodynamic limit $L \to \infty$. If $\rho$ is so that $\e^{\beta H/2} \rho \e^{\beta H/2}$ is
still bounded, then the constant $K$ is of the order $(L/\beta)^j$.

\medskip
\noindent
{\em Proof:} The proofs of both estimates are rather similar, but
the second estimate is somewhat more complicated to prove, so we only give the proof there. We
start with the operator inequalities
\ben
N_k(t) \le N_k(t)^4 \le \omega^{-2}_k N_k(t) H_0(t)^2 N_k(t) \le K \omega^{-2}_k N_k(t) H^2 N_k(t) \, ,
\een
where in the first step we used that ${\rm spec} \, N_k(t) = {\mathbb N}$, and where in the second step
we have used the obvious relations
\ben
H_0(t) =  \sum_{k \in \mz} \omega_k \, N_k(t) \ge  \omega_k N_k(t) \, .
\een
In the last step, we have used Rosen's inequality, which as we have already noted applies to
the time-translated operators as well. As usual, $K$ denotes a constant, and we
adopt the standard {\em constant convention} to denote all constants that may appear in the
various inequalities in this paper by the same letter, $K$, even though, of course, they
might be numerically different and/or depend on different parameters.
We write $\rho(A) = \tr (\rho A)$, and we apply this state to the above operator inequality, recalling
 that $\rho(N_k(t)) = n_k(t)$. We get
\bena
n_k(t) &\le& K \omega^{-2}_k \rho(N_k(t) H^2 N_k(t)) \\
&=& K \omega^{-2}_k \Bigg(
\rho(\delta(N_k(t))^* \delta(N_k(t))) + \rho(H N_k(t)^2 H) + 2 \re \, \rho (H N_k(t) \delta(N_k(t)))
\Bigg) \, . \non
\eena
We now introduce the shorthand $X(t) = \delta(N_k(t)) (I + N(t))^{-\phalf}$, with $p$ the degree of
the interaction polynomial and $N(t)$ the total number operator at time $t$. With this shorthand, we then
have the estimate
\ben
\rho\Big( \delta(N_k(t))^* \delta(N_k(t)) \Big) \le \| X(t) \|^2 \rho ((I+N(t))^{p}) \le K \| X(t) \|^2 \rho((I+H)^{p}) \, .
\een
In the first step, we used that $\rho(B AA^* B^*) \le \|A\|^2 \rho(BB^*)$, and in the second step we are
using Rosen's inequality. Using the Cauchy-Schwarz inequality and the Rosen estimate, we also have
\ben
\begin{split}
\Big| \re \, \rho (H N_k(t) \delta(N_k(t))) \Big| &\le \Big( \rho(H (I+N(t))^{\frac{p}{2}} N_k(t)^2 (I+N(t))^{\frac{p}{2}} H) \Big)^{\half} \Big( \rho (X(t) X(t)^*) \Big)^{\half} \\
&\le \c \omega_k^{-1} \| X(t) \| \Big( \rho(H (I+N(t))^{\frac{p}{2}} H_0(t)^2 (I+N(t))^{\frac{p}{2}} H)\Big)^{\half} \\
&\le \c \omega_k^{-1} \| X(t) \| \Big( \rho ((I+H)^i) \Big)^{\half} \, ,
\end{split}
\een
for some $i$.
In the second line, we have used that $N$ and $H_0$ commute and that $N_k \le \omega_k^{-1} H_0$.
We also have
\ben
\rho(H N_k(t)^2 H) \le \omega_k^{-2} \rho(H H_0(t)^2 H) \le K \omega_k^{-2} \rho(H^i) \, ,
\een
for some $i$,
again using Rosen's inequality.
Using now the assumption that $\rho(H^j) < \infty$ for
any $j$, we have altogether shown that
\ben
n_k(t) \le K \omega_{k}^{-2}(\omega_k^{-1} + \|X(t)\|)^2 \, .
\een
Thus the proof is complete if we can demonstrate the following
\begin{lemma}
For each $\epsilon>0$ there is a constant $K$ so that
\ben
\|X(t)\| = \| [N_k(t), H](I + N(t))^{-\phalf} \| \le K \omega_k^{-1+\epsilon} \, .
\een
\end{lemma}
{\em Proof:}
We calculate that $[N_k(t), H] = \lambda [N_k(t), V(t)]$ is a finite sum of operators $W$ of the form
\ben
W(t) = \omega_k^{-\half} \sum_{q_1, \dots, q_n \in \mz} \frac{\delta(k-q_1)\delta(\sum_{i \in X} q_i- \sum_{l \in Y} q_l )}{\prod_{i \neq j} \omega_{q_i}^{\half}} \prod_{i \in X} a_{q_i}(t)^* \prod_{l \in Y} a_{q_l}(t) \, ,
\een
where $n \le p$, and where $X \cup Y = \{1, \dots, n\}$. By Prop.~1.2.3 on p.~21 of \cite{gj}, we have
\ben
\| W(t) (I+N(t))^{-\phalf} \| \le \omega_k^{-\half} \left\| \frac{\delta(k-q_1)\delta(\sum_{i \in X} q_i- \sum_{l \in Y} q_l )}{\prod_{i \neq j} \omega_{q_i}^{\half}} \right\|_{\ell^2}  \, ,
\een
where on the right side we mean the $\ell^2$ norm of a function in the variables $q_i \in \mz$. We are now going to
show that $\| \dots \|_{\ell^2} \le \c \omega^{-\half+\epsilon}$, which implies the statement of the lemma. We prove this estimate
for simplicity of notation in the case when $Y = \emptyset, j=1$. Then the $\ell^2$ norm is
\ben
\| \dots \|_{\ell^2}^2 = \sum_{q_1, \dots, q_n \in \mz} \frac{1}{\omega_{k-q_1} \omega_{q_1-q_2} \dots \omega_{q_{n-1}-q_n}} \le
\c \omega_k^{-1+\epsilon} \, ,
\een
the estimation of which can be reduced successively to the estimate $\sum_{q \in \mz} \omega_{k-q}^{-1} \omega_{q}^{-1+\delta}
\le \c \omega^{-1+2\delta}_k$, where $\delta$ is small and positive. To show the last estimate, we can argue e.g. as follows
for large $|k|$:
\bena
\sum_{q \in \mz} \omega_{k-q}^{-1} \omega_{q}^{-1+\delta} &=& \left( \sum_{|q| \le \half |k|} + \sum_{|q|> \half |k|} \right)
\omega_{k-q}^{-1} \omega_{q}^{-1+\delta} \non\\
&\le& K \left( |k|^{-1+\delta} \log |k| + |k|^{-1+2\delta} \sum_{q \in \mz} \omega_q^{-2\delta} \omega_{k-q}^{-1+\delta} \right) \non\\
&\le& K  |k|^{-1+2\delta} \left( 1 +  \sum_{q \in \mz} \omega_q^{-1-\delta} \right)  \le K \omega^{-1+2\delta}_k
\, ,
\eena
where in the last line we have used H\"older's inequality and that $\log x \le K x^{\delta}$ for large $x \ge 0$ and
$\delta > 0$.
\qed

\medskip
\noindent
Later, we will also consider a perturbative expansion of the quantities $n_k(t,\lambda)$ in the coupling
constant $\lambda$. Such perturbation expansions are known not to converge, but it is still of interest to
know to what extent they can be trusted as asymptotic series. Unfortunately, we have been unable to get reliable
estimates on quantities like the remainder term in the perturbative expansion up to a given order. But it is
possible to get, without too much difficulty, estimates on related quantities, e.g. if we let the time parameter
$t$ be imaginary. The same type of arguments also provide estimates on the error in the perturbation
expansion of $n_k(t, \lambda)$ for small $t$, essentially because the function in question is analytic in $t$ for suitable
states $\rho$.

Despite the fact that our estimates on the error term in the perturbation expansion are not satisfactory for the main
purpose of this paper, the development of the Boltzmann equation, we nevertheless present our arguments here, since
they provide, at least moral, support of the  use of the perturbation expansion, and they are also maybe of use
as a general illustration of our method, which should be applicable also in other contexts. We first describe the states
that we consider, which are the density matrix states of the form $(\beta > 0)$
\ben
\rho = \e^{-\beta H/2} \sigma \e^{-\beta H/2} \, .
\een
We are allowing both $\rho$ and $\sigma = \sigma^*$ to depend on $\lambda$, and we postulate that
\ben
\left\| \frac{d^n}{d\lambda^n} \sigma \right\| \le \c^n
\een
for some constant and all $n=0,1,2,\dots$. Under this assumption, $n_k(t)$ can be continued
analytically to complex $t$ as long as $\im \, t < \beta/2$. Our result in particular covers the case $\sigma = I$, i.e.
$\rho = \e^{-\beta H}$. Our result is now the following:
\begin{thm}
Let $\rho_\lambda$ be a density matrix for each $\lambda \ge 0$
satisfying the hypothesis above for some $\beta>0$, and let $n_k(t,\lambda) = \tr ( \rho_\lambda \e^{itH(\lambda)} N_k \e^{-itH(\lambda)} )$. Let $r_N(t)$ be the remainder in be the Taylor expansion
of $n_k(it)$ up to order $N$. Then for $t < \beta/2$, we have the estimate
\ben
|r_N| \le \frac{(\lambda \c)^N (Np)!}{N!\omega_k} \, .
\een
\end{thm}
Proof: We write $n_k(it,\lambda)$ as a Taylor series to $N$-th order. The remainder in this series is given by the
Schlomilch formula
\ben
r_{N}(\lambda, t) = \frac{\lambda^N}{N!} \Int_0^1 (1-s)^{N} ( \partial^N_\lambda n_k ) (it, s\lambda) \, ds \, .
\een
Thus, we have to estimate the $N$-th $\lambda$-derivative
\ben
\frac{d^N}{d\lambda^N} n_k(it,\lambda) = \frac{d^N}{d\lambda^N} \tr \left( \sigma \e^{-(\beta/2-t)H} N_k \e^{-(\beta/2 +t)H} \right) \, .
\een
When we carry out these derivatives, they get distributed over the factors inside the trace. When derivatives
hit $\sigma$, then we use our assumption that this is estimated by the factor $\c$ raised to the number of derivatives.
When $j$ derivatives hit one of the exponential factors, we use the iterated Duhamel formula
\ben
B_j = \frac{d^j}{d\lambda^j} \e^{-s_0 H} = \Int_{s_0>s_1>...>s_j>0} \e^{-(s_0-s_1)H} V \e^{-(s_1-s_2)H} V \cdots \e^{-(s_{j-1}-s_j)H} V \e^{-s_jH} \, d^j s \, .
\een
We will use this for $s_0 = \beta/2-t>0$.
Using the inequalities $\tr(AB) \le \| A\| \tr |B|$ and $|\tr(AB)| \le (\tr |A|^2)^{\half} (\tr |B|^2)^{\half}$, it is straightforward to see that the desired estimate will
follow if we can show that
\ben
\left( \tr | B_j N_k^{\half} |^2 \right)^{\half} \le \frac{\c^j (jp)!}{\omega_k^{\half}} \, .
\een
Using that $N_k \le \omega_k^{-1} H_0 \le \c \omega_{k}^{-1} H$ by Rosen's inequality, it hence suffices to show
\ben\label{integr}
\Int_{s_0>s_1>...>s_j>0} \Bigg(\tr |
\e^{-(s_0-s_1)H} V \e^{-(s_1-s_2)H} V \cdots V \e^{-s_jH} H^{\half} |^2 \Bigg)^{\half} \, d^j s
\le \c^j (jp)!  \, .
\een
In order to continue, it is convenient to work in the
$\QQ$-space representation. For $1 \le q \le \infty$, we introduce as usual the H\"older space $L^q(\QQ,d\nu)$
as the Banach space with the norm $\|\Psi\|_q = (\int |\Psi|^q \, d\nu)^{1/q}$, noting that $\H = L^2$.
In these spaces we have the usual H\"older inequality which states that $|(\Phi,\Psi)| \le \| \Phi \|_p \| \Psi \|_q$,
when $1 = \frac{1}{p} + \frac{1}{q}$. This implies in particular
that $\| \Psi \|_r \le \| \Psi \|_q$ if $q \ge p$, since $d\nu$ is a probability measure of total weight 1.
If $T: L^r \to L^q$ is a linear operator, then we let $\|T\|_{q,r}$ be the operator norm; the ordinary
operator norm on $\H = L^2$ is a special case of this.
The H\" older norm of $V$ can be estimated as follows for any $j$, see e.g. Lem.~2.1.6 on p.~30 of~\cite{gj}:
\ben
\begin{split}
\| V \|_{2j} &= \left( \int_{\mathcal Q} |V(q)|^{2j} \, d\nu \right)^{1/2j} = (V^j \Omega_0, V^j \Omega_0)^{1/2j} \\
&\vspace{0.2cm}\\
&\le (\| V(N+I)^{-\frac{p}{2}} \| \cdot \| (N+I)^{\frac{p}{2}} V^{j-1} \Omega_0 \|)^{1/j} \\
&\vspace{0.2cm}\\
&\le \c (1+(j-1)p/2)^{p/2j} \|V^{j-1}\Omega_0\|^{1/j} \, .
\end{split}
\een
In the last step, we use (see e.g. Prop.~1.2.3 and~1.2.5 of~\cite{gj}) $\|V(N+I)^{-p}\| \le \c$, as well as the fact that the $V^{j-1} \Omega_0$ contains
at most $p(j-1)$ ``particles''. An induction then shows
\ben
\| V \|_{2j} \le \c  (j!)^{p/2j} \le \c  j^{p/2} \, ,
\een
where in the last step we have used Stirling's formula. Thus, as a multiplication operator,
$V$ is an operator $L^{r} \to L^{q}$ for $q = r-1/(\c j)$ with norm $(\c j)^p$. The second ingredient is the fact (see e.g.
Thm.~2.2.5 on p.~36 of~\cite{gj}) that
$\e^{-sH}$ is a contraction from $L^{r}$ to $L^{q}$
long as $\frac{1}{q} \le (1+ \c s)\frac{1}{r}$. Therefore (by duality),
\ben
\| \e^{-sH} \|_{q,r} \le 1 \, , \quad \text{if} \quad \frac{1}{q} \le (1+\c s)\frac{1}{r} \, , \quad
\text{and} \quad
\left( 1- \frac{1}{q} \right) \le (1 + \c s) \left( 1- \frac{1}{r} \right)\, .\non
\een
These facts are now put together to estimate the integrand of eq.~\eqref{integr}.

We first note that of the $j$ interval lengths $s_0 - s_1, \dots, s_{j-1} - s_j, s_j$, at least one will
be greater than or equal to $s_0/j$. We consider two cases. Case (1) occurs if the interval in question is
the last one, i.e. when $s_j>\tau/j$. Case (2) covers the rest. Both cases are dealt with in a similar fashion,
so we will for brevity only deal with, say, case (2).
Let $i$ be the interval in question, $s_i - s_{i+1} > s_0/j$. We first estimate
$$\e^{-s_j H} H^{\half} \le K s_j^{-\half} \, , $$
which leads
to
\bena
&& \tr |
\e^{-(s_0-s_1)H} V \e^{-(s_1-s_2)H} V \cdots V \e^{-s_jH} H^{\half} |^2 \non\\
&\le& \c s_j^{-1}  \tr  \left| \prod_{k=1}^j \e^{-(s_{k-1}-s_k)H} V \right|^2 \non\\
&\le& \c s_j^{-1} \tr (\e^{-(s_i-s_{i+1}) H} V) \left\| XX^* V\e^{-(s_i-s_{i+1})H} Y^*Y \right\| \, ,
\eena
where we are using the shorthand notations
$$X  = \prod_{k=0}^{i-1} \e^{-(s_k-s_{k+1})H} V, \qquad
Y = \prod_{k=i+1}^{j-1} \e^{-(s_k-s_{k+1})H} V \, .$$
The trace term on the right side is now estimated using that $s_i - s_{i+1} > s_0/j$.
In order to tame the factor of $V$ under the trace, we write $V = V(I+N)^{-p}(I+N)^p$, and we use the
Rosen inequality to estimate $(I +N)^{p} \le \c H^{p}$, as well as the identity (see e.g.
Props.~1.2.3 and~1.2.5 of \cite{gj}) $\|V(I+N)^{-p}\| \le \c$. This gives
\ben
\tr (\e^{-(s_i-s_{i+1}) H} V) \le \| \e^{-s_0 H/(2j)} V\| \cdot \tr (\e^{-s_0 H/(2j)})
\le \e^{\c j/s_0} \| H^p \e^{-s_0 H/(2j)} \| \le \e^{\c j/s_0} \, ,
\een
with a constant $\c$. In order to estimate $\left\| XX^* V\e^{-(s_i-s_{i+1})H} Y^*Y\right\|$,
we use the mapping properties of the multiplication operator $V$ and the contractions $\e^{-\tau H}$ between
the H\"older spaces $L^p$. We then get
\ben\label{XY}
\left\| XX^* V\e^{-(s_i-s_{i+1})H} Y^*Y\right\| \le \prod_{k=1}^{2j+1} (\c j)^p \le [\c^j (jp)!]^2
\een
because each of the $2j+1$ factors of $V$ has norm $(\c j)^p$ as an operator from $L^r \to L^q$,
where $q = r - 1/(\c j)$. This decrease in the H\"older index for each factor of $V$ is compensated by
the increase in the H\"older index caused by the $2j+1$ heat kernels $\e^{-(s_k-s_{k+1})H}$ in the $X$ and $Y$
in eq.~\eqref{XY}. In the last step we have used Stirling's formula. Thus, in total we have shown that in
case (2), we have
\ben
\Bigg(\tr |
\e^{-(s_0-s_1)H} V \e^{-(s_1-s_2)H} V \cdots V \e^{-s_jH} H^{\half} |^2 \Bigg)^{\half} \le s_j^{-\half} \c^j (jp)!
\een
and upon integration over the $s_k$, this gives the desired bound eq.~\eqref{integr}. Case (1) is dealt with similarly. \qed

\section{Pre-Boltzmann equation for the $\phi^{p}$-model in two dimensions}
\label{prebolz}
\subsection{Derivation}

In this section we combine the estimates obtained for the quantum field model of the previous section~\ref{sec3}
and the projection technique recalled in section~\ref{sec2} in order to obtain a preliminary
form of the Boltzmann equation [see eq.~\eqref{preboltzmann} below] for the expected number densities $n_k(t) = \tr[\rho N_k(t)]$.
This equation is an exact equation which
holds non-perturbatively, and we will refer to it as the ``pre-Boltzmann equation''. It has some key features in
common with the Boltzmann equation that we will eventually derive, but it differs from the latter also
in some ways, in particular, the latter is not an exact equation, and it only holds in the thermodynamic limit $L \to \infty$,
and the long-time limit, $t \to \infty$. In this section, we will not yet take these limits.

We first need to define the reference states $w_t$ (compare eq.~\eqref{wtdef})
for our model and the set of observables $N_k$ where $k \in \mz$. In accordance with our constructions in sec.~\ref{sec2},
we let $w_t$ be the density matrix state
\ben
w_t(X) = \tr(\rho_t X) \, , \qquad \rho_t = Z(t)^{-1} \, \exp \left( -\sum_{k \in \mz} \mu_k(t) N_k(t)  \right) \, ,
\een
where $X$ is e.g. a bounded operator, and where the quantities $\mu_k$ are defined through the formula
\ben\label{mukdef}
\mu_k(t) = -\log \frac{n_k(t)}{n_k(t)+1} > 0 \, .
\een
To see that this formula makes sense, let us assume first $n_k(t) > 0$ for all $k \in \mz$. As shown in thm.~\ref{nestim}
in the previous section~\ref{sec3}, when the initial state  density matrix state $\rho \in {\mathscr I}_1(\H)$ is such
that also $\rho H^j \in {\mathscr I}_1(\H)$ for all $j$, then we have $n_k(t) \le \c \omega_k^{-4+\epsilon}$. It then easily follows
that the state sum $Z(t) = \prod_{k \in \mz} (1-\e^{-\mu_k(t)})^{-1}$ is convergent:
\ben
\begin{split}
Z(t) &= \exp \Big( - \sum_{k \in \mz} \log(1-\e^{-\mu_k(t)}) \Big) \\
& \le \exp \Big( K \sum_{k \in \mz} \e^{-\mu_k(t)} \Big) \\
& \le \exp \Big( K \sum_{k \in \mz} \omega_k^{-4+\epsilon} \Big) < \infty \, .
\end{split}
\een
Thus, $\rho_t$ is indeed a trace class operator for any $t \in \mr$, and the state $w_t$ is well-defined.
The situation is the same when some $n_k(t) = 0$, essentially because this means that $\mu_k(t) = +\infty$, and this
only improves the convergence properties. Thus, $w_t$ is a well-defined state if the initial state of the system $\rho$
is such that $\rho H^j$ has a finite trace for all $j \ge 0$.

A different way to characterize the state $w_t$ is to say that it is the unique quasifree state
w.r.t to the time $t$-observables $a^\#_k(t)$, whose 2-point function is
\ben\label{wtaa}
w_t(a_k(t)^* a_j(t)) = n_k(t) \, \delta(k-j) \, ,
\een
and whose $n$-point functions are zero for an odd number of creation/annihilation operators,
and factorize into 2-point functions for an even number. More precisely,
for $X,Y \subset \mz$
\ben\label{wtwick}
w_t \left( \prod_{k \in X} a_k(t)^* \prod_{j \in Y} a_j(t) \right) =
\delta_{|X|,|Y|} \sum_{f: X \to Y \,\,{\rm bijective} } \prod_{j \in X} w_t(a_j(t)^* a_{f(j)}(t)) \, .
\een
It is important to realize that for fixed $t$, this factorization formula for $w_t$ will not hold for
the creation and annihilation operators $a_k(s)$ at another time $s \neq t$ unless the model is free, $\lambda = 0$.
The above factorization formula also demonstrates once again that the state is well-defined also when one or more $n_k(t)$'s
happen to be equal to zero, and we have
\ben\label{nkt1}
n_k(t) = w_t(N_k(t)) \, , \qquad \text{for all $t \in \mr$, $k \in \mz$.}
\een
The covariance matrix is found to be diagonal,
\ben
c_{jk}(t) = w_t [(N_j(t) - n_j(t)I)(N_k(t)-n_k(t)I)] = \frac{1}{L} n_k(t)(n_k(t) + 1) \, \delta(k-j) \, .
\een
We now define our projector according to the general recipe laid out in section~\ref{sec2}. To be on the safe side,
we first consider only a subset $\{N_k \mid k \in \mz, |k| \le \Lambda\}$ of observables where $\Lambda < \infty$,
and we put $\mu_k = + \infty$ for $|k| > \Lambda$ in $w_t$ and denote the correspondingly
changed state as $w_t^\Lambda$. This change has the effect that eq.~\eqref{nkt1} is valid only for
$|k| \le \Lambda$, and that eq.~\eqref{wtaa} returns zero for $|k|>\Lambda$.
The projector
as in eq.~\eqref{projdef} is then:
\ben\label{projectorx}
\begin{split}
\P_t^\Lambda(A) &=
w_t^\Lambda(A) \, I + \sum_{|j| \le \Lambda} \frac{(N_j(t)-n_j(t)I) \cdot w_t^\Lambda((N_j(t) - n_j(t)I) A)}{n_j(t)(n_j(t) + 1)} \\
& = w_t^\Lambda(A) \, I + \sum_{|j| \le \Lambda} (N_j(t)-n_j(t)I) \, \frac{\partial}{\partial n_j(t)} w_t^\Lambda(A)
\, .
\end{split}
\een
To arrive at the formula in the second line, we used the analogue of eq.~\eqref{trick} for $A_j=N_j$.
In writing that expression, we have also anticipated that, in the expressions below, $A$ will be a power series in
$a_k^\#(t)$, and $w_t(A)$ can then be written as a corresponding power series in $n_k(t)$ by eq.~\eqref{wtwick}.
The operator $\partial/\partial n_k(t)$ then acts as the usual partial derivative operator on such an expression.
In particular, it is clear from the last line that the $n_j(t)$'s in the denominator in the first line will
always cancel, and so the case $n_j(t) = 0$ will never cause any problems.

We also need to choose our initial conditions, i.e. the quantum state $\rho$ that we would like to
investigate. As we have just explained, in order for the states $w_t$ to be well defined, we require
$\rho H^j$ to have finite trace for sufficiently large non-negative $j$. Furthermore, we would
like to have an initial state so that the Robertson equation is valid without a ``memory term'',
compare sec.~\ref{sec2}, thm.~\ref{robertson}. Thus, we would ideally like to choose as our state
as $\rho = \rho_0$, where
\ben\label{initialcond}
\rho_0 = \frac{1}{Z} \exp \left( -\sum_{k \in \mz} \mu_k N_k \right) \, .
\een
for some $\mu_k$. In other words, we would like to choose our state to be quasi-free (w.r.t. to the time-0
creation/annihilation operators $a^\#_k(0)$!), and we would also like our initial state to be translation
invariant\footnote{That is, invariant under the 1-parameter group generated by the momentum operator
$P= \Sum_k (k/L) \, a_k^* a_k$.}.
In the finite dimensional context, we were free to make this assumption. Unfortunately, in the present model
with infinitely many degrees of freedom,
a technical difficulty arises because we also need the initial state $\rho$ to satisfy the condition
that $\rho H^j$ be trace class for sufficiently large $j$. This condition is needed not only in order to guarantee
that the $w_t$ are well defined for all times $t \in \mr$, but it turns out to be essential also in order
to give sense to the other ingredients in the Robertson equation the present infinite dimensional context, see
below. Unfortunately, there seems to be a conflict between demanding that $\rho H^j \in {\mathscr I}_1(\H)$,
and that $\rho$ be quasifree, i.e., equal to~\eqref{initialcond} for some $\mu_k$. The reason for
this conflict seems to be the presence of the non-trivial interaction $\lambda V$ in the Hamiltonian.
The problem disappears if we only interpret the pre-Boltzmann equation in the perturbative sense (see sec.~\ref{sec5}),
but here we wish to have formulae that hold non-perturbatively. So we are forced to introduce e.g. a damping factor
into eq.~\eqref{initialcond}, such as taking\footnote{In order to see that this state satisfies
$H^j \rho \in {\mathscr I}_1$, one can e.g. use the Rosen estimates, noting that on the right side
we can replace $H$ by $H + \sum_k \mu_k N_k$.}
\ben
\rho = Z^{-1} \, \exp \left( -\beta H -\sum_{k \in \mz} \mu_k N_k \right)
\een
where $\beta$ is arbitrarily small but positive [compare thm.~\ref{trace}]. This has the effect of creating a memory term
in the Robertson equation, whose form is given by eq.~\eqref{rob1}, with $A_j = N_j$. The memory
term will clearly be of order $\beta$. We will not bother much about the memory term,
since this depends on the precise choice of initial state. Also, when we pass to the perturbative
expansion and long-time-dilute-medium limits in the next sections, we can take the initial state as quasifree, and in that case the memory term vanishes.

\medskip

With our definition of the projection operators etc. in place, we can now formally appeal to the result obtained above in
thm.~\ref{robertson}, see eq.~\eqref{rob1}. This gives us:

\begin{thm} (``Pre-Boltzmann equation'')
Let $|p| \le \Lambda$, where $\Lambda < \infty$, and let the state $\rho$ satisfy the ``initial condition''~\eqref{initialcond}. Then we have the following equation for the expected number densities $n_p(t) = \tr(\rho N_p(t))$, dropping
the reference to $\Lambda$ on $w_t$:
\bena\label{preboltzmann}
\frac{1}{\lambda^2} \frac{d}{dt} n_p(t) &=& \text{``memory term''} - \Int_0^t ds \,\, w_s^\Lambda ([V(s),[V(t),N_p(t)]])  \hspace{2mm}
- \\
&& \sum_{r=1}^\infty (-i\lambda)^r \Int_0^t ds \,\,  \sum_{|k_1|, \dots, |k_r| \le \Lambda} \,\, \Int_{t>\sigma_r >...>\sigma_{1}>s}
w_{s}^\Lambda ([V(s),[V(\sigma_1),N_{k_1}(\sigma_1)]]) \cdot \non\\
&&
\prod_{j=1}^{r}
\frac{\partial}{\partial n_{k_j}(\sigma_j) } w_{\sigma_j}^\Lambda ([V(\sigma_{j+1}), N_{k_{j+1}}(\sigma_{j+1})])
\, d^r \sigma \non
\eena
where $k_{r+1} = p, \sigma_{r+1} = t$ in the last factor in the product. The infinite sum on the right side converges absolutely.
The derivatives $\partial/\partial n_k(\sigma_j)$ are understood as explained below eq.~\eqref{projectorx}.
The ``memory term'' is given by $\tr[(\rho-\rho_0) \E_{0,t}(\delta N_p(t))]$, and it can be expanded in a similar absolutely
convergent series.
\end{thm}

\medskip
\noindent
{\bf Remark}: Before we come to the proof of this theorem, we emphasize that the sum over $r$ in formula
given in the theorem is {\em not}
a perturbation series in $\lambda$, which is known {\em not} to converge. This is because
the order $m$ term contains terms that are themselves functions of $\lambda$, for example
a term like $w_s([N_k(t), V(t)])$ is a function of $\lambda$, the Taylor series for which would
not converge. Also, $\Lambda$ is {\em not} a cutoff of the theory, but merely a restriction on
the set of momenta $k$ in $n_k(t)$ that we monitor.

\medskip
\noindent
{\em Proof:} We formally take $A_j = N_j$ in thm.~\ref{robertson}, which gives
\begin{equation}\label{eq:eqn_motion_decomposed2}
\frac{d}{dt} n_p(t) = \text{``memory term''} -\Int_{0}^t
          w_s \left[ \delta  \circ \E_{s,t} \circ \delta(N_p(t)) \right] \, ds \, .
\end{equation}
Then, if we formally substitute the series expression~\eqref{eseries} for $\E_{t,s}$
with our choice~\eqref{projectorx} of projectors $\P_t^\Lambda, t \in \mr$, then we
arrive at the expression given in the theorem after a few simple manipulations.
Of course this does not conclude the proof, because equation~\eqref{rob0}
was originally derived only in the context of matrices and it is not a priori
clear to what extent it makes sense in the infinite dimensional context considered now.
The main question is whether the cocycle $\E_{s,t}$ can be defined in the
infinite dimensional setting, and the second question is to what extent the
above compositions make sense, i.e. whether the domains match up.

We first establish that the evolution cocycle $\E_{t,s}$ is well defined. We define it
as above by the series expression~\eqref{eseries}, but of course we cannot use
the proof given there to show that the series is also convergent in the present setting. Instead,
we will need to give a new proof, the result of which we state as a
\begin{lemma}\label{lem4}
The series for $\E_{s,t}(A)$ converges for any $A$ with finite ``Sobolev''-norm $\| A \|_1 = \|(I+H)^{-1} A (I+H)^{-1}\|$,
and we have in fact
\ben\label{etsest}
\|\E_{s,t}(A)\|_1 \le \e^{O(\lambda)|t-s|} \| A \|_1 \, .
\een
In other words $\E_{s,t} : H B(\H) H \to H B(\H) H$ is a bounded operator with
exponentially bounded norm on the closure $H B(\H) H$ of $B(\H)$ under $\| \, . \, \|_1$.
\end{lemma}

The lemma proves that the domains in the composition
$\E_{t_1,t_2} \circ \E_{t_2,t_3}$ match up, and the series formula~\eqref{eseries} for the evolution
cocycle then also shows that the cocycle condition indeed holds,
\ben
\E_{t_1,t_2} \circ \E_{t_2,t_3}(A) = \E_{t_1,t_3}(A) \quad \text{for all $A \in H B(\H) H$.}
\een
From this, it is now simple to demonstrate that it satisfies the desired differential
equation. Let us prove the lemma.

{\em Proof of lemma~\ref{lem4}:} Let $A$ be a bounded operator. From the series expression for $\E_{t,s}$, we can estimate, dropping the superscript ``$\Lambda$'' on the projectors and $w_t$ for simplicity:
\ben\label{series1}
\begin{split}
&\|\E_{s,t}(A)\|_1 \\
&\le \sum_{k=0}^\infty \Int_{t>\sigma_k>...>\sigma_1>s}
\|\delta \circ \P_{\sigma_1}  \circ \cdots
\delta \circ \P_{\sigma_k}(A) \|_1 \, d^k\sigma \\
& \le \sum_{k=0}^\infty \lambda^k \Int_{t>\sigma_1>...>\sigma_k>s} d^k \sigma \sum_{|j_1|,...,|j_k| \le \Lambda}
\| [H, N_{j_1}(\sigma_1)] \|_1 \,  \, \Big| w_{\sigma_k}\Big( (N_{j_k}(\sigma_l) - n_{j_k}(\sigma_l)I) A \Big) \Big| \\
& \hspace{1cm} \cdot \prod_{l=1}^{k-1} \frac{1}{n_{j_l}(\sigma_l)} \,
\Big|
w_{\sigma_l}\Big( (N_{j_l}(\sigma_l) - n_{j_l}(\sigma_l)I)
[V(\sigma_{l+1}), N_{j_{l+1}}(\sigma_{l+1})] \Big)
\Big| \, .
\end{split}
\een
Our aim is to show that each term under the sum on the right side
can be bounded by $\| A \|_1 (|t-s| \lambda \c \Lambda)^k/k!$,
which will imply that the series converges absolutely for any $t$, and the
inequality~\eqref{etsest}.

In order to estimate the integrand, we use another lemma, which is
at the heart of our analysis, and which makes crucial use of the estimates
$n_k(t) \le K \omega_k^{-4+\epsilon}$ that were derived above in thm.~\eqref{nestim}.
\begin{lemma}\label{lem5}
For all $k \in \mz$ we have
\ben\label{first}
|w_s([N_k(t), V(t)] N_j(s))| \le \c\,  n_j(s) \, \omega_j \omega_k^{-1}
\een
as well as
\ben\label{second}
|w_s([V(s),[V(t), N_k(t)]])| \le \c \,  \omega_k^{-1}
\een
uniformly in $t$ and $s$.
Furthermore, if $A$ is a bounded operator, we also have
\ben\label{third}
|w_s(N_j(s) A)| \le \c \omega_j n_j(s) \| A \|_1
\, .
\een
\end{lemma}
A proof of this lemma is given below.
When we now use the estimates~\eqref{first} and~\eqref{third} from the lemma on the terms under the last integral, we
note that the dangerous factors of $n_j(s)^{-1}$ precisely cancel out with the corresponding factor in the estimate,
and the subsequent factors of $\omega_j$ and its inverse also cancel.
In formulae, we have
\bena
&& \| \delta \circ \P_{\sigma_1}  \circ \cdots
\delta \circ \P_{\sigma_k}(A)  \|_1 \non \\
&\le& (K\lambda)^k \, \| A \|_1 \sum_{|j_1|, \dots, |j_k| \le \Lambda} \| [H, N_{j_1}(\sigma_1)]  \|_1 \,
\omega_{j_k} n_{j_k}(\sigma_k)
\, \prod_{l=2}^k
\frac{n_{j_{l-1}}(\sigma_{l-1}) \, \omega_{j_{l-1}}^{} \omega_{j_l}^{-1}}{n_{j_{l-1}}(\sigma_{l-1})}
 \non\\
&\le& (\c \lambda \Lambda)^k \, \|  A  \|_1 \, .
\eena
In the third line we have used that for an interaction polynomial of degree $p$, we have
using $N_k(t) \le \omega_k^{-1} H_0(t)$, and Rosen's inequality:
\ben
\begin{split}
\|[H,N_j(t)]\|_1 &= \| R [H,N_j(t)] R \| \\
&\le 2 \| R N_j(t)^2 R \|^{\half}\\
& \le 2 \omega_j^{-\half + \epsilon} \| R H_0(t)^{1-2\epsilon} N(t)^{1+2\epsilon} R\|^{\half}\\
& \le \c \omega_j^{-\half + \epsilon} \| R H^2 R\|^{\half} \le \c \omega_j^{-\half + \epsilon} \, ,
\end{split}
\een
with $R = (I+H)^{-1}$.
When $p \le 4$, 
we can improve this result to
$\| R [H,N_j(t)] R \| \le \c \omega_j^{-1}$
using an improved version of Rosen's inequality holding in that case, see e.g. sec.~3.1 of~\cite{gj}. This is what has been used above
for definiteness, but the case of general $p$ is analogous and leads only to slightly worse estimates.
If we now take into account that the volume of the integration region is $|t-s|^k k!^{-1}$, we get
\bena\label{series2}
\|\E_{s,t}(A)\|_1
&\le& \sum_{k=0}^\infty \Int_{t>\sigma_k>...>\sigma_1>s}
\|\delta \circ \P_{\sigma_1}  \circ \cdots
\delta \circ \P_{\sigma_k}(A) \|_1 \, \, d^k\sigma \, \non\\
&\le& \sum_{k=0}^\infty \frac{(\c|t-s| \lambda\Lambda)^k}{k!}  \| A \|_1 = \e^{O(\lambda)|t-s|} \|A\|_1 \, ,
\eena
so the convergence of
the series~\eqref{series1} follows for all $t$ for any $A$ with
$\| R A R \| < \infty$.  This proves lemma~\ref{lem4}. \qed

To complete the proof of the theorem, the only further thing
we need to check is that the combination $w_s[\delta \circ \E_{t,s} \circ \delta (N_p(t))]$ is
well defined. We can estimate  $w_s[\delta \circ \E_{t,s} \circ \delta (N_p(t))]$ writing down again the series expression as above,
using the same type of argument as just given, and using also eq.~\eqref{second}. Then we see that the $k$-th term in the
sum is now dominated by $(\c \Lambda \lambda |t-s|)^k /k!$, where the $|t-s|^k k!^{-1}$ again
comes from the volume of the set $\{t>\sigma_k>\dots>\sigma_1>s\}$. This shows convergence, and
completes the demonstration of the theorem up that of the above lemma~\ref{lem5}.\qed

\medskip
\noindent
{\em Proof of lemma~\ref{lem5}:}
The proof of all estimates is rather similar; we show the first estimate~\eqref{first}. We have
\bena
w_s([N_k(t), V(t)] N_j(s)) &=& w_s([N_k(t), V(s)] N_j(s)) \non\\
&=& w_s(N_k(t) V(s) N_j(s)) - w_s(N_j(s) V(s) N_k(t)) \, .
\eena
The terms on the right side are estimated in exactly the same manner. We demonstrate the argument for one of them.
We have, using the Cauchy-Schwarz inequality together with the fact that $w_s(N_k(t) V(s) N_j(s)) = w_s(N_j(s)^{\half} N_k(t) V(s) N_j(s)^{\half})$:
\ben
\Big|w_s(N_k(t) V(s) N_j(s)) \Big| \le \Big( w_s(N_j(s)^{\half} N_k(t)^2 N_j(s)^{\half} ) \Big)^{\half} \Big( w_s(N_j(s)^{\half} V(s)^2 N_j(s)^{\half}) \Big)^{\half} \, .
\een
By Rosen's inequality, together with the inequality
$(\Psi, (A+B)^2 \Psi) \le 2 (\Psi, (A^2 + B^2)\Psi)$ for hermitian $A,B$,
we can estimate
$$N_k(t)^2 \le \c \omega_k^{-2} H^2 \le 2 \c \omega_k^{-2} (H_0(s)^2 + \lambda^2 V(s)^2),$$
so we find
\ben
\Big| w_s(N_k(t) V(s) N_j(s)) \Big|^2 \le \c \omega_k^{-2} \cdot w_s \Big( N_j(s) (H_0(s)^2 + \lambda^2 V(s)^2) \Big) \,
w_s \Big( N_j(s) V(s)^2 \Big) \, . \non
\een
We continue our estimation by using Prop.~1.2.3 of \cite{gj}, which implies that
$$V(s)^2 \le \| V(s)(I+N(s))^{-\phalf}\|^2 \, (I+N(s))^{p} \le \c (I+N(s))^{p}.$$
This then allows us to estimate
\ben
\Big| w_s(N_k(t) V(s) N_j(s)) \Big| \le
\c\omega_k^{-1}  \, \left\{ w_s \Big(N_j(s) (I+N(s))^{p} \Big)^2
+  w_s\Big(H_0(s)^2 N_j(s) \Big)^2 \right\}^{\frac{1}{2}} \,
\, .
\een
Now, we recall that the total number operator is $N = \sum_{k\in \mz} N_k$, and the free Hamiltonian
is $H_0 = \sum_{k \in \mz} \omega_k N_k$, and similarly for the quantities at time $s$. Thus, in view of the inequality
just given, we have reduced the problem to that of estimating quantities of the form $w_s(N_{k_1}(s) \dots N_{k_r}(s))$.
For this, we need a simple combinatorial formula. To derive this formula, let $X$ be any finite subset of $\mz$, and for
each $i \in X$, let $\alpha_i \in {\mathbb N}$. Then we have, using elementary Fock-space algebra:
\ben
\begin{split}
w_s \left( \prod_{i \in X} N_i(s)^{\alpha_i} \right) &= \prod_{i \in X} \frac{\partial^{\alpha_i}}{\partial \xi_i^{\alpha_i}} \,\,
w_s \left( \exp \sum_{i \in X} \xi_i N_i(s) \right) \Bigg|_{\xi_i = 0}\\
&=\prod_{i \in X} \frac{\partial^{\alpha_i}}{\partial \xi_i^{\alpha_i}} \,\,
\frac{1-\e^{-\mu_i(s)}}{1-\e^{-\mu_i(s)+\xi_i}} \Bigg|_{\xi_i =0}\\
&=\prod_{i \in X} \frac{\partial^{\alpha_i}}{\partial \xi_i^{\alpha_i}} \,\,
\frac{1}{1+(1-\e^{\xi_i})n_i(s)} \Bigg|_{\xi_i =0} =\prod_{i \in X} n_i(s) \psi_{\alpha_i}(n_i(s)) \, .
\end{split}
\een
Here $\psi_n(x)$ are the degree $(n-1)$ polynomials defined iteratively
\ben
\psi_1(x) = 1 \, , \qquad \text{and} \qquad
\psi_{n+1}(x) = (1+x) \frac{d}{dx} [x \psi_n(x)] \, .
\een
This formula implies (dropping the reference to ``$s$'' in $n_i(s)$ on the right side to lighten the notation):
\ben\label{wscomb}
\begin{split}
w_s\left( N_{k_1}(s) \cdots N_{k_p}(s) \right)
&= \Sum_{X_1 \cup ... \cup X_n = \{1, ...,p\}}
\sum_{l_1,...,l_n \in \mz} \prod_{j \in X_1} \delta(l_1-k_j) \cdots \prod_{j \in X_n} \delta(l_n-k_j) \\
&\hspace{2cm}
\cdot \,\,
n_{l_1} \psi_{|X_1|}(n_{l_1}) \cdots n_{l_n} \psi_{|X_n|}(n_{l_n}) \, .
\end{split}
\een
With the help of this formula, we can now easily estimate quantities like e.g. $w_s(N(s)^p)$. In
such an expression we have a $p$-fold iterated sum over expressions of the form~\eqref{wscomb}.
The key point is now that, after taking into account the Kronecker delta's, we are left with iterated sums each of which
is accompanied by at least one factor of $n_k(s)$. Because we have the estimate $n_k(s) \le \c \omega_k^{-4+\epsilon}$ from thm.~\ref{nestim},
such a sum will converge. If we have e.g. an expression of the form $w_s(H_0(s)^2 N(s)^p)$, we can make a similar argument.
Now, after taking into account the Kronecker delta's, we are left with iterated sums which, at worst contain
a factor of $\omega_k^2$ (from the squared free Hamiltonian), and at least one factor of $n_k(s)$. Again,
because we have the estimate $n_k(s) \le \c \omega_k^{-4+\epsilon}$ from thm.~\ref{nestim},
such a sum will converge. By making simple arguments of this kind, we thus easily arrive at the basic estimates:
\ben\label{upthere}
w_s \left( N_j(s) H_0(s)^2 \right) \le \c n_j(s) \omega_j^2 \, , \qquad
w_s \left( N_j(s) (I+N(s))^{p} \right) \le \c n_j(s) \,,
\een
from which it follows that $|w_s(N_k(t) V(s) N_j(s))| \le \c n_j(s) \omega_k^{-1} \omega_j$, and
we find the same estimate for $|w_s(N_j(s) V(s) N_k(t))|$. This concludes the proof of the first inequality~\eqref{first}.

The second inequality~\eqref{second} is dealt with in a very similar fashion. For the third inequality, we can argue
e.g. by saying that
\bena
w_s(N_k(s) A) &=& w_s (N_k(s)^{\half} A N_k(s)^{\half} ) \non\\
&\le& \| (I+H)^{-1} A (I+H)^{-1} \| \,
w_s (N_k(s)^{\half} H^2 N_k(s)^{\half} ) \, \, .
\eena
We continue the estimation by
\ben
\begin{split}
& w_s (N_k(s)^{\half} H^2 N_k(s)^{\half} ) = w_s (N_k(s) H^2  )\\
&\vspace{0.1cm}\\
&\le 2 w_s(N_k(s) H_0(s)^2) + 2 \lambda^2 w_s(N_k(s) V(s)^2) \\
&\vspace{0.1cm}\\
&\le 2 w_s(N_k(s) H_0(s)^2) + 2 K \lambda^2 w_s(N_k(s) (I+N(s))^p) \, .
\end{split}
\een
The expressions on the right side have already been estimated in eq.~\eqref{upthere}, and
hence the desired inequality~\eqref{third} follows.
\qed

\medskip
\noindent

\subsection{Alternative form of the pre-Boltzmann equation}

It is convenient for later purposes to write the pre-Boltzmann equation in a way which makes more manifest the
dependence of the integrands on the number densities $n_j(s)$. For this, it is convenient, to introduce the ``collision kernels'' $B^\Lambda(E,p,s)$ by the
formula
\begin{equation}\label{eq:B_def}
B^\Lambda(E,p,s) := \frac{-i \lambda E}{2\pi}
 \Int_{\mr} dt~ \Exp^{-i E(t-s)}
w_s^\Lambda \left( [V(s), N_p(t)] \right) \, .
\end{equation}
From now on we will drop the reference to $\Lambda$, for simplicity of notation.
The terminology for these kernels will become later below, where we will relate them to scattering cross sections.
These kernels are distributions in $E$ that are defined for any $s \in \mr$ and $p \in \mz$. We claim that the pre-Boltzmann
equation can be written entirely in terms of these kernels. We will demonstrate this now for the collision term on the
right side of the Boltzmann equation. Similar arguments can also be applied to the memory term. However, this will
later be set to zero anyway by an appropriate choice of initial state, so we will not discuss this here.

The statement is clear for the first term on the right side of the
pre-Boltzmann equation, since the factor of $E$ in front of $B(E,p,s)$ can be converted to a $t$-derivative
in the integrand, which in turn yields the resulting first term on the right side of the pre-Boltzmann equation
in view of $\frac{\partial}{\partial t} N_p(t) = i[H, N_p(t)] = i\lambda[V(t),N_p(t)]$, using $[H_0(t), N_p(t)]=0$.
The remaining terms on the right side of the pre-Boltzmann equation can also be massaged into convolutions
of the collision kernel using the following elementary chain of equalities:
\begin{equation}\label{eq:Rescattering_one}
\begin{split}
\lambda \, w_\tau( [V(t),N_p(t) ]) &=
w_\tau ([H_0(t) + \lambda V(t),N_p(t)]) =
w_\tau ([H(t),N_p(t)])  = \\
&\vspace{4mm}\\
& = w_\tau( [H(\tau),N_p(t)])  =
\lambda \, w_\tau( [V(\tau),N_p(t)])   = \\
&\vspace{4mm}\\
& = \lambda \, w_\tau( [V(\tau),N_p(t)])  -
\lambda \, w_\tau( [V(\tau),N_p(\tau)])   = \\
& \hspace{-2cm} = \lambda \Int_{\tau}^t d\tau'~
w_\tau ( [V(\tau), \partial_{\tau'}N_p(\tau')]) 
=i\lambda^2 \Int_{\tau}^t d\tau'~
w_\tau ( [V(\tau),[V(\tau'),N_p(\tau')]] )~.
\end{split}
\end{equation}
This equality puts the terms appearing in the second half of the pre-Boltzmann equation into a form
similar to the first, and so we can again express them through the collision kernel $B(E,p,s)$.
In the above computation we have used $[H_0(t), N_p(t)] = 0$ in the first and fourth equality. To go from the
first line to the second line we use energy conservation, i.e.~the fact that the full Hamiltonian
does not depend on time. Combining eq.~\eqref{eq:Rescattering_one}, the definition of $w_t$,
and that of the collision factor, eq.~\eqref{eq:B_def} then leads to the following proposition:

\begin{prop}
The pre-Boltzmann equation can be expressed in terms of the collision kernels $B(E,p,s)$ as
\begin{equation}\label{eq:preBeqn}
\begin{split}
\frac{d}{dt} n_p(t) &= \text{``memory term''} - \Int_{0}^t d s \Int_{\mr} d E~\Exp^{iE(t-s)} B(E,p,s) \hspace{5mm} - \\
& \sum_{n=1}^\infty 
\Int_{0}^t ds \Int_{\mr} dE~
\Int_{\Delta_{2n}(s,t)}  d^n \tau d^n \sigma \Int_{\mr^n} d^n E \negfive
\sum_{|k_1|,\ldots,|k_n| \le \Lambda} \cdot \\
&\hspace{-.3cm}\cdot \Exp^{iE(\tau_1 - s)} B(E,k_1,s)
%
\prod_{j=1}^n \Exp^{i E_j(\sigma_j-\tau_j)} \frac{\partial}{\partial n_{k_j}\!(\tau_j)} B(E_j,k_{j+1},\tau_j)
\, .
%
\end{split}
\end{equation}
In this expression, we are denoting by $\Delta_{2n}(s,t) = \{s<\tau_1<\sigma_1<\dots<\tau_n<\sigma_n<t\}$,
and $k_{n+1} = p$ in the expression under the integral. This is an equivalent form of the pre-Boltzmann
equation, and hence still valid non-perturbatively. As above, the sum over $n$ is absolutely convergent.
\end{prop}

The collision kernels $B(E,p,s)$ that appear in this form of the pre-Boltzmann equation can be rewritten
in terms of (local) M\" oller operators and the number densities $n_j(s), j \in \mz$, as we now explain.
Define the local M\" oller operators as
\ben
S(s,t) := \e^{isH} \e^{-i(s-t)H_0} \e^{-itH} = \e^{i(t-s)H_0(s)} \e^{-i(t-s)H} \, .
\een
These are unitary operators on $\H$. The true $S$-matrix of the theory (which exists for the model, see~\cite{hk})
would be given in terms of these local M\" oller operators by $S = \slim_{t \to \infty} S(0,+t)S(-t,0) =
\slim_{t \to \infty} S(-t,t)$
on a suitable domain of vectors in $\H$, in the infinite volume limit $L \to \infty$. However, we
will not take these limits here as yet. We also record the properties
\ben\label{ptsst}
S(s,s) = I \, , \qquad
\frac{\partial}{\partial t} S(s,t) \Phi = (-i\lambda) \, \e^{i(t-s)H_0(s)} V(s) \e^{-i(t-s)H_0(s)} \,  S(s,t)  \Phi \, ,
\een
for suitable vectors $\Phi \in \H$. The terminology ``local M\" oller operators'' arises from the fact that they
are equal to the $S$-matrix of a theory wherein the interaction $V$ is switched only within the time-interval $[s,t]$.
We will come back to this when we look at the perturbative expansion for the local M\" oller operators in the next
section.

From the properties of $S(s,t)$, together with the fact that
$\e^{i(t-s)H} N_p(s) \e^{-i(t-s)H} = N_p(t)$, we see that
\bena\label{commut1}
\lambda \, w_s([V(s), N_p(t)]) &=& -i \,  \frac{\partial}{\partial t} \, w_s \left( \e^{i(t-s)H} N_p(s) \e^{-i(t-s)H}  \right) \non\\
&=& -i \, \frac{\partial}{\partial t} \, w_s \left( S(s,t)^* N_p(s) \, S(s,t) \right) \,  \\
&=& -i \, \frac{\partial}{\partial t} \, w_s \left( S(s,t)^* [N_p(s), S(s,t)] \right)
\, .\non
\eena
We can expand the local M\" oller operators in terms of creation and annihilation operators as
\ben\label{Sst}
S(s,t) = I + i\Sum_{X,Y \subset \mz} {\mathcal M}_{X \to Y}(s,t) \frac{\delta(k_X-k_Y)}{\prod_{i \in X} (2\omega_i)^{\half}
\prod_{j \in Y} (2 \omega_j)^{\half}}
\prod_{j \in Y} a_j(s)^* \prod_{i \in X} a_i(s) \, .
\een
The delta is due to the fact that the local M\" oller operators commute with the momentum operator on $\H$,
and $k_X$ denotes the sum over all $i \in X$ etc.
Then, combining this formula with that for the
collision kernels $B(E,p,s)$, we get
\bena
&&B(E,p,s) \non \\
&=& \,\, \frac{E^2}{2\pi L} \Sum_{X,X',Y,Y'} \,\,  \Int_\mr dt \,\, \e^{-iE(t-s)} \, (\delta_X(p) - \delta_Y(p)) \,
{\mathcal M}_{X \to Y}(s,t) \overline{{\mathcal M}_{X' \to Y'}(s,t)} \,  \\
&& \delta(k_X-k_Y) \delta(k_{X'}-k_{Y'}) \, w_s \left(
\prod_{i' \in X'} \frac{a_{i'}(s)^*}{(2\omega_{i'})^{\half}} \prod_{j' \in Y'} \frac{a_{j'}(s)}{(2\omega_{j'})^{\half}}
\prod_{j \in Y} \frac{a_j(s)^*}{(2\omega_j)^{\half}} \prod_{i \in X} \frac{a_i(s)}{(2\omega_i)^{\half}}
\right) . \non
\eena
The summation is over finite subsets $X,Y,X',Y'$ of $\{-\Lambda, \dots, \Lambda\} \subset \mz$, and we use the notation
\ben
\delta_X(p) =
\begin{cases}
L & \text{if $p \in X$,}\\
0 & \text{otherwise.}
\end{cases}
\een
The expectation values are directly evaluated using ``Wick's theorem'' in the form of
formula~\eqref{wtwick}. Applying this formula gives us, after some combinatorial
considerations, the expression:
\ben\label{B1}
\begin{split}
& B(E,p,s) = \frac{1}{2\pi}
\sum_{X,Y}' \, (\delta_X(p) - \delta_Y(p)) \,  \Int_\mr dt \,\, \e^{-iE(t-s)} \\
& \hspace{1cm}
\cdot \,\,\, E^2 |\widetilde{\mathcal M}_{X \to Y}(s,t)|^2  \,  \delta(k_X-k_Y)
\prod_{i \in X} \frac{n_i(s)}{2\omega_i} \prod_{j \in Y} \frac{1+n_j(s)}{2\omega_j} \,,
\end{split}
\een
where the sum is over all subsets $X,Y \subset \{-\Lambda, \dots, \Lambda\} \subset \mz$, and where the ``dressed''
matrix elements (with tilde) are defined by
\ben\label{dressed}
\widetilde{\mathcal M}_{X \to Y}(s,t) = \sum_{Z \subset \{-\Lambda,...,\Lambda\}}' {\mathcal M}_{X \cup Z \to Y \cup Z}(s,t)
\prod_{i \in Z} \frac{n_i(s)}{2\omega_i} \, .
\een
The dressed matrix elements have an expansion in terms of Feynman diagrams, but
the propagators are not the standard Feynman propagators, but instead are
modified by contributions depending on $n_i(s), i \in \mz$. These expressions
will be given in the next section.
We may summarize our findings in this section in the following proposition:

\begin{prop}
The expected number densities $n_j(t)$ satisfy the integro-differential equation~\eqref{eq:preBeqn},
where the collision kernels $B(E,p,s)$ are given by eq.~\eqref{B1}. These kernels depend on the
number densities as well as the local scattering matrix elements ${\mathcal M}_{X \to Y}$ via
eq.~\eqref{dressed}.
\end{prop}

\noindent
\medskip
{\bf Remark:} In the next section we will see that the dressed matrix elements $\widetilde{\M}_{X \to Y}$
have an expansion in terms of Feynman diagrams with ``dressed'' propagators.

\section{Perturbative expansion}\label{sec5}

In the preceding section, we have derived an integro-differential equation (see eq.~\eqref{eq:preBeqn}) for the expected number
densities $n_j(t)$. This equation is hardly simpler than the original Heisenberg equation of motion
for the corresponding operator quantities $N_j(t)$,
but it has the advantage that it is not an operator equation, and that it only involves
the quantities $n_j(t)$, and no other information.  What we will do is to take the
collision factors $B(E,p,s)$ (see eq.~\eqref{B1}) in the series, and expand each one of them in a perturbation series in $\lambda$.
In this way, we will relate the expressions under the integral signs in the pre-Boltzmann equation
to standard integrals (position space Feynman integrals) in perturbation theory.
Thus, we will get a closed system of equations for the collection $n_j(s)$ in terms of quantities that
are in principle calculable in perturbation theory. This will important for the next
sections when we derive the Boltzmann equation.

We will from now on ignore all questions related to the convergence of the perturbation series, which at any rate
is known not to converge. Thus, in this section, all power series in $\lambda$ that we will write down are to be understood
as {\em formal} power series only---note that this was not the case for the series in the pre-Boltzmann equation,
which we proved to be convergent. If we neglect such questions, we are then free to choose the initial state $\rho$
to be quasifree, i.e. of the form $\rho_0$ in eq.~\eqref{initialcond}, and this has the effect that the
``memory term'' disappears in eq.~\eqref{eq:preBeqn}. It is then also possible to take $\Lambda = \infty$,
to get a closed system of equation simultaneously for all number densities, and we will do this from now on.
Apart from this, we do not make any simplification, in particular, we will not drop
any terms in the perturbation expansion.

As we have said, the terms that we would like to express in perturbation theory are the
terms under the integral in the pre-Boltzmann equation. As seen from the pre-Boltzmann equation in the form~\eqref{preboltzmann}, these terms have a
relatively simple expression in terms of the time-$s$-creation/annihilation operators $a_k^{\#}(s)$ [see eq.~\eqref{akt}], and so it is natural to attempt a perturbation expansion  about a fictitious free field $\phi_0(t,x)$ that coincides with the interacting field $\phi(t,x)$ at time $s$---rather than an asymptotic in-field as is more commonly done in perturbation theory.
The free field $\phi_0(t,x)$ that we will consider is an operator-valued
distribution on $\H$, satisfying the free massive Klein-Gordon equation in the sense of distributions
\ben
(\square - m^2) \phi_0(t,x) = 0 \, ,
\een
with $\square$ the Klein-Gordon operator on the flat Lorentzian cylinder $\mr \times S^1$ of circumference
$2\pi L$. This free field $\phi_0(t,x)$ is taken to satisfy the same initial conditions at time $t=s$ [when we
take a matrix element with any
$\Phi, \Psi \in {\mathscr D}(H^{\half})$, and in the sense of distributions in $x \in S^1$] as the interacting field $\phi(t,x)$:
\ben
\phi_0(t,x)  = \phi(t,x)  \, , \qquad \text{and} \quad
\tfrac{\partial}{\partial t} \phi_0(t,x)  = \tfrac{\partial}{\partial t} \phi(t,x)  \, , \quad
\text{when $t=s$.}
\een
Note that these conditions only hold for the time $t=s$, and will be false for other times $t \neq s$, unless
the coupling constant $\lambda$ is zero. Indeed, the interacting fields $\phi$ evolves with respect to the
full hamiltonian $H$, whereas the free field $\phi_0$ evolves with respect to the time-$s$ free Hamiltonian
$H_0(s)$. Note also that the free field $\phi_0$ depends on the initial time $s$ at which it is set to be equal to
the interacting field. Strictly speaking, this should be incorporated into the notation $\phi_0(t,x)$ somehow,
but we will not do this here for simplicity. The free field has a simple expression in terms of the
creation and annihilation operators $a^{\# }_k(s)$, which is given by
\ben\label{phi0}
\phi_0(t,x) = \Sum_{k \in \mz} [u_k(t,x) \, a_k(s)^* + \overline{u_k(t,x)} \, a_k(s) ] \, ,
\een
where $u_k(t,x)$ are the standard positive frequency solutions to the free Klein-Gordon equation on
$\mr \times S^1$ given above in eq.~\eqref{ukt}.  As is well known, the interacting field $\phi$ can be
expressed in terms of the free field via a formal power series expression.
This series is usually given in the case that the free field $\phi_0$ is an ``in''-field (i.e. formally taking $s \to -\infty$), and it is then called ``Haag's series''~\cite{haag,glz}. But a similar formula is also valid for finite $s$, see e.g.~\cite{bf,hw3,df}; it is given below.

After these preparations,
we are now ready to give a perturbative expansion for the collision kernels
$B(E,p,s)$, see eq.~\eqref{eq:B_def}. There are actually two such representations, and we will give them both
in the following two subsections. The first one is in terms
of the local M\" oller operators and their perturbative expansion (see previous section), whereas the second
one is in terms of retarded products. The first derivation has the advantage that it directly involves the (local)
scattering cross sections, and it will be used in the next section when we come to the long-time limit.
The second expression is more suitable when working in curved spacetime, and this will be the starting point
of our investigation in the second paper~\cite{hl2} in this series. Furthermore, the second derivation also
goes through in higher spacetime dimensions $d>2$ when renormalization becomes an issue, whereas the first
derivation is only valid as it stands in superrenormalizable models.

\subsection{Expression 1}
The first way is to
use equation eqs.~\eqref{Sst},~\eqref{B1} which contain the local M\" oller operators $S(s,t)$, where $s<t$, and their matrix elements. These operators
satisfy the differential equations~\eqref{ptsst}, which, when written in terms our auxiliary free field read
\ben
S(s,s) = I \, , \qquad
\frac{\partial}{\partial t} S(s,t) \Phi = -i\lambda \, V_0(t) S(s,t)  \Phi \, .
\een
Here, we have introduced the ``interacting picture'' potentials, given in terms of the auxiliary free field by
\ben
V_0(t) = \sum_{n=0}^p b_n \, \Int_0^{2\pi L} :\phi_0^n(t,x): \,\, dx \, ,
\een
and we have used that $V_0(t) = \e^{i(t-s)H_0(s)} V(s) \e^{-i(t-s)H_0(s)}$, because
$H_0(s)$ generates the time evolution of the free field $\phi_0(t,x)$.
The double dots in the above equation mean that we expand the free field in terms of $a_k^{\#}(s)$ and
move all annihilation operators to the right, in formulae
\ben
:\phi_0^n(t,x):
\,\,\, = \Sum_{|X\cup Y| =n} \,\,\,
\prod_{k \in X} u_{k}(t,x) a_{k}(s)^*  \prod_{j \in Y} \overline{u_{j}(t,x)} a_{j}(s) \, , \non
\een
 where the sum runs over subsets $X,Y \subset \mz$.
The differential equations are readily integrated in the sense of formal power series in $\lambda$, and this gives
\ben
S(s,t) = \sum_{r=0}^\infty (-i\lambda)^r \Int_{s<\sigma_1<...<\sigma_r<t} V_0(\sigma_r) \cdots
V_0(\sigma_1) \, d^r \sigma \, .
\een
The right side is a time ordered exponential. It is clear from this expression that $S(s,t)$ agrees with the ``local
$S$-matrix'' of Bogoliubov~\cite{b} and Epstein-Glaser~\cite{eg}.
We now expand the time ordered products into normal ordered products of the free field $\phi_0$ using an appropriate
version of Wick's theorem (see e.g. the ``local Wick expansion'' of \cite{hw1,bf}). Then we obtain a perturbative
expression for the matrix elements ${\mathcal M}_{X \to Y}(s,t)$ [see eq.~\eqref{Sst}] in terms of position space
Feynman integrals with a ``time cutoff'' restricting the integration range of the time variables of the Feynman
integrals to the interval $[s,t]$. Tn the present superrenormalizable model these integrals are absolutely convergent, without the need of any sort of renormalization process beyond the normal ordering procedure which has already been carried out.

For completeness, we give the formula here. For a Feynman graph $G$, with interaction vertices as given
by the interaction polynomial in eq.~\eqref{Hdef}, let $V(G)$ be the set of vertices,
$L(G)$ the set of internal lines, $E(G)$ the set of vertices connected to external lines. For each subsets $X,Y \subset \mz$ of
momenta, and $j \in E(G)$, let $X(j) \subset X$ be the ingoing momenta from $X$ connected to that vertex $j$, each associated
with an external line, and similarly we let $Y(j) \subset Y$ be the outgoing momenta connected to $j$. The perturbative
expansion for the matrix elements of the local M\" oller operators is then
\bena\label{mxypert}
{\mathcal M}_{X \to Y}(t,s) &=& \sum_{r=0}^\infty (-i\lambda)^r \sum_{G: |V(G)|=r} c_G \, (2\pi)^{-\half |X| - \half |Y|}
\Int_{s < x_1^0,...,x_r^0 < t}
d^{2r} x \non\\
&& \prod_{ij \in L(G)} \Delta_F(x_{i},x_j) \prod_{j \in E(G)} 
\exp \left(-i k_{X(j)} x_j + i k_{Y(j)} x_j \right)
\eena
Here, $x_{i} = (x^0_i, x^1_i) \in \mr \times S^1$ are spacetime points, and $k_X$ is the sum of all
2-momenta $(k_0,k_1) =(\omega_k,k/L)$ corresponding to a set $X \subset \mz$ etc. $c_G=|{\rm Aut}(G)|^{-1} \prod_{i \in V(G)} b_{n(i)}$
is a standard weight factor associated with the diagram (with $n(i)$ the valence of the $i$-th vertex) which is
explained in more detail in many textbooks on quantum field theory. The $\Delta_F$ are the Feynman propagators---i.e. time ordered vacuum expectation values of free fields $\phi_0$---of the cylinder spacetime
$\mr \times S^1$, given by
\ben\label{Feynmanp}
\begin{split}
\Delta_F(x_1,t_1,x_2,t_2) &= \frac{1}{2\pi L} \sum_{p \in \mz} \Int_{\mr} dE \, \frac{\e^{-iE(t_1-t_2)+ip(x_1-x_2)/L}}{
\omega_p^2 - E^2 - i0} \\
&= \frac{1}{4\pi} \sum_{n \in \mz} K_0(\sqrt{\sigma_n + i0})
\, ,\\
&\vspace{3mm}\\
\sigma_n & = m^2[-(t_1-t_2)^2 + (x_1 - x_2 + 2\pi n L)^2] \, .
\end{split}
\een
To go from the first to the second line, we have used the
the Poisson summation formula, and $K_0$ denotes a Bessel function. The last formula
says the that Feynman propagator on the cylinder arises from that on Minkowski spacetime
by taking a ``sum over images''. The relevant point for us here is that $K_0(\sqrt{z}) \sim \log z$,
so the singularities of the Feynman propagator on the cylinder will only be present for
null-related pairs of points, and
are locally $L^p$-functions for any $p<\infty$, thus implying the absolute convergence of the integral in formula~\eqref{mxypert}.

A simple combinatorial argument then shows that the ``dressed'' matrix elements
$\widetilde{\mathcal M}_{X \to Y}(s,t)$ [see eq.~\eqref{dressed}] are  given by the same formula, but with the ``dressed'' propagators
\ben\label{tildelta}
\begin{split}
\widetilde \Delta_F(x_1,t_1,x_2,t_2) &= \theta(t_1-t_2) w_s(\phi_0(x_1,t_1) \phi_0(x_2,t_2)) +
\theta(t_2-t_1) w_s(\phi_0(x_2,t_2) \phi_0(x_1,t_1)) \\
&= \Delta_F(x_1,t_1, x_2,t_2) + \frac{1}{2\pi L} \sum_{k \in \mz} \frac{n_k(s)}{2\omega_k} \e^{-i\omega_k (t_1-t_2) + ik(x_1-x_2)/L}
\, .
\end{split}
\een
In order to see that the integrals for the dressed propagators are still absolutely convergent, it
is sufficient to show that the term involving the sum is a sufficiently regular function. In fact it is of class
$C^{3} (\mr \times S^1 \times \mr \times S^1)$. This easily follows from the fact
that $n_k(s) \le \c \omega_k^{-4+\epsilon}$ for any $\epsilon>0$, by thm.~\ref{nestim}.

\medskip
\noindent
Thus, {\em in summary,} with our choice for the initial state $\tr(\rho \, . \,) = w_0$ (``maximum entropy''), the memory term on the right side of the
pre-Boltzmann equation~\eqref{eq:preBeqn} vanishes. The
collision kernels $B(E,p,s)$ in the pre-Boltzmann equation are still given by the above expression, which we repeat:
\ben
\begin{split}
& B(E,p,s) = \frac{1}{2\pi}
\Sum_{X,Y} \, (\delta_X(p) - \delta_Y(p)) \,  \Int_\mr dt \,\, \e^{-iE(t-s)} \non\\
& \hspace{1cm}
E^2 |\widetilde{\mathcal M}_{X \to Y}(s,t)|^2  \,  \delta(k_X-k_Y)
\prod_{i \in X} \frac{n_i(s)}{2\omega_i} \prod_{j \in Y} \frac{1+n_j(s)}{2\omega_j} \,,
\end{split}
\een
where $X,Y$ run over subsets of $\{-\Lambda, \dots, \Lambda\} \subset \mz$.
These are in turn obtained from the dressed matrix elements $\widetilde{\mathcal M}_{X \to Y}(t,s)$, see eq.~\eqref{dressed}.
We have seen that the following proposition is true:

\begin{prop}
The perturbative expansion (i.e. formal power series expansion in $\lambda$) of the dressed matrix
elements is:
\ben\label{tilMpert}
\begin{split}
\widetilde{{\mathcal M}}_{X \to Y}(t,s) &= \sum_{r=0}^\infty (-i\lambda)^r \sum_{G: |V(G)|=r} c_G \, (2\pi)^{-\half |X| - \half |Y|} \Int_{s < x_1^0,...,x_r^0 < t}
d^{2r} x \\
& \hspace{0.3cm} \prod_{ij \in L(G)} \widetilde \Delta_F(x_{i},x_j) \prod_{j \in E(G)} 
\exp \left(-i k_{X(j)} x_j + i k_{Y(j)} x_j \right) \, ,
\end{split}
\een
where $\widetilde \Delta_F$ is the dressed Feynman propagator on $S^1 \times \mr$ given by eqs.~\eqref{tildelta},~\eqref{Feynmanp}. The integrals converge absolutely.
\end{prop}
This concludes our first derivation of the perturbative expansion for the right side of the
pre-Boltzmann equation.

\medskip
\noindent

\subsection{Expression 2}
We now give an equivalent alternative way to expand $B(E,p,s)$ in a formal power series in $\lambda$.
This derivation relies on an alternative expansion of the interacting field $\phi$ in terms of the free field $\phi_0$.
The relevant formula is well-known in the literature (``Haag's series''), see e.g.~\cite{haag,bf, df,steinmann}. It is, in our notation
\ben
\begin{split}
&\phi(t,x) = \phi_0(t,x) \hspace{1cm} + \\
& \hspace{1cm} \sum_{n=1}^\infty \frac{(i\lambda)^n}{n!} \, \Int_{[s,t]^n} \R_n \Big( \phi_0(t,x); V_0(\tau_1)
\otimes \cdots \otimes V_0(\tau_n) \Big) \, d^n \tau \quad
\text{for $t \ge s$.}
\end{split}
\een
Here, the notation $\R_n$ means a ``retarded product''. Mathematically, it is convenient to take the view
that it is a bilinear map $\R_n: {\mathcal F} \times (\otimes^n {\mathcal F}) \to {\rm End}(\H)$, where $\F$ is the
linear space of {\em classical} local expressions of a fictitious classical field $\phi$ of the form
$A =\int W[\phi_0(x), \partial \phi_0(x), \dots, \partial^r \phi_0(x)] f(x) \, d^2 x$, with $W$ a multivariate polynomial,
and $f \in C^\infty(S^1 \times \mr)$. In
particular, in $\F$, no field equations are assumed. The $\R_n$ take their values in a suitable space of quadratic
forms in $\H$, for details see~\cite{hw1,hw2,bf,df}. The retarded products are distributional in nature, i.e.
for $B,A_1,\dots,A_{n} \in \F$ of the above form, $\R_n(B;\otimes_i A_i)$ is a distribution
in the test functions $h,f_1, \dots, f_n$ implicit in $B,A_1,\dots,A_n$. For example, inside
the retarded product, $V_0(t) = \sum b_n \int \phi_0(t,x)^n \, dx$
means the {\em classical} expression for the potential (hence no ``normal ordering''), and it is not understood that the classical field
$\phi_0$ is to satisfy a field equation when standing inside $\R_n$.  For a single factor and $W(x) = W[\phi_0(x), \partial \phi_0(x), \dots, \partial^r \phi_0(x)]$, we have
\ben
\R_0(W(x)) = \,\,\,  :W(x): \, .
\een
Thus, the formula for the interacting field $\phi(x)$ has
$\phi_0(x)$ as its lowest order term, as required. If $B,A_i$ are smeared polynomials $U,W_i$ in $\phi_0(x)$ but {\em not}
its partial derivatives (viewed again as ``classical expressions''), then there is a similarly simple expression also
for the corresponding retarded product $\R_n(B; \otimes_i A_i)$ with $n$ factors; it is given by a sum of multiple commutators multiplied by step functions, see e.g.~\cite{df}.
However, if the arguments of the retarded products contain derivatives, then this simple formula
becomes ill-defined, essentially because one then has to perform renormalization\footnote{When carrying out this renormalization,
it turns out to be of considerable advantage to consider the arguments of the retarded products to be classical expressions,
and this is why we proceed here in this way.} In this case, the retarded products may be thought of as defined by
a combinatorial formula in terms of
time-ordered and anti-time ordered products $T$ resp. $\overline T$ (see e.g.~\cite{bf,df}), which is
\ben
\R_n\Big(B; \bigotimes_{i=1}^n A_i\Big) = \sum_{X \cup Y = \{1, \dots, n\} } \bar T_{|X|+1}\Big( B \otimes \bigotimes_{j \in X} A_j \Big) \, T_{|Y|}\Big( \bigotimes_{i \in Y} A_i \Big) \, .
\een
This then leaves one with the product of defining the ordinary time ordered products, see e.g.~\cite{eg,bf,df,hw1,hw2}.
It follows from these constructions that the time ordered/retarded products
have an expression in terms of $a_k^\#(s)$, the creation/annihilation operators at time $s$, cf.~\eqref{akt}.

The retarded products owe their name to their support properties: If $B,A_i$ are polynomials in $\phi_0$ and
its partial derivatives, viewed as ``classical expressions'', then we have
\ben
\supp \R_n\Big(B; \otimes_{i=1}^n A_i\Big) \subset \{ \supp A_i \subset J^+( \supp B) \quad \text{for all $i=1, \dots, n$} \} \, ,
\een
where $J^\pm(S)$ denotes the causal future/past of a set $S \subset \mr \times S^1$, and
where we define the support of an expression $A =\int W[\phi_0(x), \partial \phi_0(x), \dots, \partial^r \phi_0(x)] f(x) \, d^2 x$ to
be equal to the support of the testfunction $f$.

We can now start with our task of expanding expressions on the right side of the pre-Boltzmann equation
in perturbation theory. This
is accomplished essentially by inserting a perturbative formula for the interacting field $N_p(t)$
into eq.~\eqref{eq:B_def}. In order to do this in an efficient way, we proceed as follows. First, we note that,
by the Glaser-Lehmann-Zimmermann (GLZ)-formula (see \cite{df,glz}), we have\footnote{Note that $N_p(t)$ when
expressed in terms of the free field $\phi_0$ is not a local expression in $\F$. However, it is still local in time,
and this is sufficient in order for the retarded (or time ordered) product to make sense in two spacetime dimensions.}
:
\bena
&&[V(s), N_p(t)] = \\
&&\sum_{n,m = 0}^\infty \frac{(-i\lambda)^{n+m}}{(n+m)!}
\Int_{[s,t]^{n+m}} \left[ {\mathcal R}_n \Big(V_0(s); \bigotimes_{j=1}^n V_0(\sigma_j) \Big) \, ,
{\mathcal R}_m \Big(N_p(t); \bigotimes_{l=1}^m V_0(\sigma_l) \Big) \right] \, d^{n+m} \sigma =
\non\\
&&
\sum_{n=0}^\infty \frac{(i\lambda)^n}{n!} \!\!\Int_{[s,t]^n}\!\! \left\{ \!
{\mathcal R}_n \Big(V_0(s); N_p(t) \otimes
\bigotimes_{j=1}^n V_0(\sigma_j) \Big)
-{\mathcal R}_n \Big(N_p(t); V_0(s) \otimes
\bigotimes_{j=1}^n V_0(\sigma_j) \Big) \!\! \right\} \, d^n \sigma \non \, .
\eena
Then, multiplying this equation through with a step function and using the support properties of the retarded products, we get:
\ben\label{before}
\theta(t-s) [V(s), N_p(t)] = \sum_{n=0}^\infty \frac{(i\lambda)^n}{n!} \Int_{\mr^n} {\mathcal R}_n \left(N_p(t); V_0(s) \otimes
\bigotimes_{j=1}^n V_0(\sigma_j) \right) \, d^n \sigma \, .
\een
We now use this expression in our expression~\eqref{eq:B_def} for the collision kernel $B(E,p,s)$. In this formula, we may multiply the
integrand by a step function $\theta(t-s)$, because the opposite step function $\theta(s-t)$ would give a contribution to $B(E,p,s)$
that is analytic for $\im E > 0$, and which would for this reason vanish when substituted back into the pre-Boltzmann equation. Therefore,
up to such an irrelevant contribution, our collision term becomes
\ben
\begin{split}
&B(E,p,s) = \frac{1}{2\pi} \sum_{n=0}^\infty \frac{(i\lambda)^n}{n!} \Int_{\mr} dt \,\, \e^{-iE(t-s)} \\
&\hspace{2cm}
\cdot \Int_{[s,t]^n} w_s \Big[{\mathcal R}_n \Big( N_p(t); V_0(s) \otimes
\bigotimes_{j=1}^n V_0(\sigma_j) \Big) \Big] \, d^n \sigma \, .
\end{split}
\een
This is our second expression for the collision factor.
The expectation values of the retarded products in the states $w_s$ can be evaluated in terms of Feynman integrals with
``propagators''~\eqref{tildelta} using a version of Wick's theorem~\cite{hw3,bf,df}, because the retarded products are
expressible in terms of $a_k^\#(s)$, and because the states $w_s$ are quasifree, cf.~\eqref{wtwick}. The propagators
can be evaluated in terms of the factors
$n_k(s)$, but we will not show this here. In higher dimensions $d>0$ (or for any model that is not superrenormalizable),
the fully renormalized retarded product must be understood in the above expression.

\medskip
\noindent
In summary, our pre-Boltzmann equation~\eqref{eq:preBeqn} together with the perturbative expression for $B(E,p,s)$
gives us a closed set of integro-differential equations for the unknown quantities $n_p(t)$. While these
equations are not particularly simple, but we will see in the next section that they form a good starting point for a
further expansion, namely a simultaneous expansion essentially the inverse observation time $1/t$, and the
coupling constant $\lambda$ [or typical initial density $n_p(0)$].

\section{The long-time and low-density limit}\label{longtime}

We are now in the position to study the limit $t \to \infty$ of the expected number densities $n_p(t)$.
It is in this limit that the Boltzmann equation as stated in the introduction will emerge. When taking this
limit, it is clear, however, that we must at the same time consider a correspondingly dilute medium,
or a correspondingly weakened interaction. The latter case is somewhat simpler and has been
studied previously e.g. in the context of a lattice fermi gas in~\cite{salm,hu}.
It could be discussed in our framework as well, and this would only
involve certain relatively obvious modifications of the arguments that we now present for dilute medium (low-density) limit.

The idea is to introduce a new small parameter, $\epsilon$, into the problem. This is not an additional
coupling constant, but instead characterizes the initial density matrix state $\rho(\epsilon)$ of the system, as well
as the time over which we observe it.
Roughly speaking, the time duration over which we wish to observe is of order $\epsilon^{-1}$, whereas the initial densities characterizing
the initial state via eq.~\eqref{initialcond} are of order $\epsilon^\alpha$ for a suitable $\alpha>0$. The idea is to make an expansion of the observable
quantities, i.e. the expected densities, in the new small parameter $\epsilon$. The leading order contribution in this expansion
will obey an equation closely related to the Boltzmann equation described in the introduction. But one can also consider higher orders. These will
describe corrections to the dilute-medium-and-long-time limit. In this paper, we will only consider the leading order.
The precise value of $\alpha=1$ will be needed for the limit as $\epsilon \to 0$ to exist. In order
to see more clearly how this value arises, however, we will for now we keep it as a free parameter.

\subsection{Derivation of main equation}
In order to introduce the long-time-dilute-medium limit, let us write
\ben\label{rescalings}
t = T/\epsilon \, , \quad n_p(t) = \epsilon^\alpha \,  \nu_p(\epsilon, T) \, \, .
\een
The idea is to take $\epsilon \to 0+$, while keeping $T$ fixed (so that $t \to \infty$), and while keeping the
initial density $\nu_p(\epsilon, 0)$--and hence the initial state~\eqref{initialcond}--fixed (so that $n_p(0) \to 0$). We will
also take the thermodynamic limit $L \to \infty$. We claim that
the limiting quantities
\ben
\nu_p(T) := \lim_{\epsilon \to 0} \nu_p(\epsilon, T) \, ,
\een
if they exist,
satisfy an equation [cf.~\eqref{finalmaster}] which is similar to the Boltzmann equation, and which reduces to the
Boltzmann equation as stated in the introduction if we also assume that the collision time is long
(e.g. when $\lambda$ becomes small). Because the
initial densities $n_p(0)$ are scaled to zero, the limit that we consider is the long-time-and-dilute-medium
limit.

The full mathematical
demonstration of this claim would require us to control the limit as $\epsilon \to 0$ of $\nu_p(\epsilon, T)$,
and for this we would have to look at the full non-perturbative dynamics of the model. This ought to be possible in
principle using methods similar to those described in the previous sections, but we do not believe that such an
analysis would necessarily offer considerably more insight into the nature of the limit than a formal derivation. In this
section, we would like to give such a more formal derivation. Our proof starts by simply {\em assuming}
that the limit $\nu_p(T) := \lim_{\epsilon \to 0} \nu_p(\epsilon, T)$ exists in a suitably strong sense.
We then use the pre-Boltzmann equation to see what equation this limit must satisfy. The equation that
 we arrive by this process--taking also the freedom to exchange limits and integrals while we are at it--will be eq.~\eqref{finalmaster}.

We begin by substituting $t = T/\epsilon$ into the pre-Boltzmann equation in the form~\eqref{preboltzmann}, noting
that the ``memory term'' is absent due to our choice of the initial state~\eqref{initialcond}. After
changing variables $s \to s/\epsilon$ and $E \to \epsilon E$ in the first integral and making similar
changes in the other terms, we are led to the equation
\ben\label{eq:preBeqn1}
\begin{split}
\partial_T \, \nu_p(\epsilon, T) &= - \Int_{0}^T d s \Int_{\mr} d E~\Exp^{iE(T-s)} B(\epsilon , E,p,s) \hspace{4mm} - \\
& \sum_{n=1}^\infty (-1)^n \epsilon^{n(\alpha-1)}
\Int_{0}^T d s \Int_{\mr} d E~
\Int_{\Delta_{2n}(s,T)}  d^n \tau d^n \sigma \Int_{\mr^n} d^n E \negfive
\Sum_{|k_1|,\ldots,|k_n| \le \Lambda} \cdot \\
&\hspace{-.3cm}\cdot \e^{iE(\tau_1 - s)} B(\epsilon, E,k_1,s)
\prod_{j=1}^n \e^{i E_j(\sigma_j-\tau_j)}
\frac{\partial}{\partial \nu_{k_j}\!(\tau_j, \epsilon)}
B(\epsilon, E_j,k_{j+1}, \tau_j)
\, .
\end{split}
\een
In this expression, we are denoting by $\Delta_{2n}(s,T) = \{s<\tau_1<\sigma_1<\dots<\tau_n<\sigma_n<T\}$,
and $k_{n+1} = p$ in the expression under the integral. Furthermore, we denote
\begin{equation}\label{eq:B_lim}
B(\epsilon, E, p, s) :=  \frac{1}{\epsilon^{1+\alpha}} B\left( \epsilon E,p,\frac{1}{\epsilon} s \right) ~ \, ,
\end{equation}
and it is understood that in this expression for $B$, all factors of $n_p(s)$ have been replaced by
$\epsilon^\alpha \, \nu_p(s, \epsilon)$. So far, we have only performed trivial changes of variables,
and hence our formulae still hold exactly. Now, we would like to take the limit $\epsilon \to 0+$, and
the thermodynamic limit $L \to \infty$ and the limit $\Lambda \to \infty$. As we have said, we are going to discuss these limits only in a semi-rigorous fashion,
but we believe that a rigorous, but considerably more involved, discussion would also be possible and would
lead to the same conclusion.

Let us first discuss the thermodynamic limit $L \to \infty$. Since the quantities $n_p(t)$ were defined as densities
[compare~\eqref{Nkt}], we expect that they will possess a well defined thermodynamic limit. Assuming this to be the case,
the pre-Boltzmann equation is expected to continue to hold in the thermodynamic limit, at the very least in the
sense of formal power series in $\lambda$. To obtain the collision
factors, one must then only make the (standard) replacements:
\begin{equation*}\label{eq:inf_vol_assumptions}
\begin{split}
n_p(t)~, \quad  p \in \mz \qquad &\xrightarrow{L \to \infty} \qquad
n_p(t)~, \quad p \in \mr  \\
L \, \partial/\partial n_p, \quad p \in \mz \qquad &\xrightarrow{L \to \infty} \qquad
\delta/\delta n_p~, \quad p \in \mr \\
\Int_0^{2 \pi L} dx ~
\qquad &\xrightarrow{L \to \infty} \qquad
\Int_\mr dx ~, \quad x \in \mr \\
\omega_p = \sqrt{p^2/L^2 + m^2}~, p \in \mz \qquad &\xrightarrow{L \to \infty} \qquad \omega(p) = \sqrt{p^2+m^2}~, p \in \mr\\
\frac{1}{L} \sum_{p \in \mz}  f(p/L) \qquad &\xrightarrow{L \to \infty} \qquad
\Int_\mr d p~ f(p) \\
\delta(p)~, \quad p \in \mz \quad \text{(scaled Kronecker delta)} \qquad &\xrightarrow{L \to \infty} \qquad
\delta(p)~, \quad p \in \mr \quad \text{(Dirac delta)}.
\end{split}
\end{equation*}
The functional derivative in the second line
is defined as usual by $\frac{d}{dz} F[n+zf] |_{z=0}= \int \delta F [n]/\delta n_p \, f_p \, dp$.

In order to write the rescaled collision factor~\eqref{eq:B_lim}
in the thermodynamic limit, we use that
\ben\label{CP}
{\mathcal M}_{X \to Y}(s,t) = {\mathcal M}_{Y \to X}(-t,-s)
\een
in the present PT-invariant theory, and we also use the identity
\ben\label{mxya}
{\mathcal M}_{X \to Y}(s+a,t+a) = {\mathcal M}_{X \to Y}(s,t)\, \exp \left\{-ia(\omega_X - \omega_Y) \right\} \, ,
\een
where here and in the following we set
\ben
\omega_X := \sum_{q \in X} \omega(q) \, , \quad k_X := \sum_{q \in X} q \, .
\een
Then, after a trivial change of integration variables, the
rescaled Boltzmann collision factor becomes in the thermodynamic limit (and the limit $\Lambda \to \infty$)
\ben\label{bepsform}
\begin{split}
&B\left( \epsilon, E,p,s \right) =
\frac{1}{2\pi} \Int_{X,Y} d\Pi_{X,Y} \, \delta(k_X-k_Y) \, \delta_X(p) \, \Int_\mr dt \,\, \e^{-iEt} \\
& \vspace{1.5cm} \cdot \,\,
E^2 \left| {\mathcal M}_{X \to Y}\Big(0 , \frac{t}{\epsilon} \Big) \right|^2
\Big\{\epsilon^{\alpha(|X|-1)} \prod_{q \in X} \nu_q(\epsilon, s) -  \epsilon^{\alpha(|Y|-1)} \prod_{q' \in Y} \nu_{q'}(\epsilon, s) \Big\} \,\, + \dots \,\, ,
\end{split}
\een
where the dots are higher order terms in $\epsilon$ that we have not displayed, because
they will disappear when we take $\epsilon \to 0+$.
Here $X = \{q_1, \dots, q_n\}$ resp. $Y = \{q_1', \dots, q_{n'}'\}$ now denote sets of real on-shell
momenta, $d\Pi_{X,Y}$ is the natural integration element concentrated on the corresponding number of Cartesian
copies of the upper mass hyperboloid
\ben
d\Pi_{X,Y} = \prod_{q \in X \cup Y  } d\Pi_q \, , \qquad d\Pi_q = \frac{dq}{2\sqrt{q^2+m^2}} \, ,
\een
and it is understood that $X$ resp. $Y$ run over arbitrary subsets of $\mr^{2n}$ resp. $\mr^{2n'}$,
and $n$ resp. $n'$ is summed over.

We would now like to take the limit as $\epsilon \to 0+$ of the rescaled collision factor in
eq.~\eqref{bepsform}. Let us define $\B_p(s)$ to be the expression
\ben\label{eq:Beqn_low_final0}
\begin{split}
\B_p(s) &= 4\pi^2
\Int_{{\mathbb H}(m)^{n+1}} d\Pi_{k_1,\dots,k_{n+1}}
~\cdot\\
& \Bigg\{ \delta^2\Big( p + q_1^{} -  q_2 - \dots - q_{n+1} \Big)
\Big| {\mathcal M}_{p, q_1^{} \to q_2, ..., q_{n+1}}
\Big|^2 \nu_p(s) \nu_{q_1}(s) - \\
& \hspace{-0.2cm}-
\delta^2\Big( p + q_1^{} + \dots +q_{n-1} -  q_n - q_{n+1} \Big)
\Big| {\mathcal M}_{p, q_1, ..., q_{n-1} \to q_n, q_{n+1}}
\Big|^2 \nu_{q_n} (s) \nu_{q_{n+1}}(s) \Bigg\} \, ,
\end{split}
\een
where $\M_{X \to Y}$ now denotes the {\em full} (rather than local) scattering matrix element of the theory, with
the energy-momentum conservation delta's taken out. ${\mathbb H}(m) = \{(p_0,p_1) \in \mr^2 \mid
p_0 = \sqrt{p_1^2+m^2} \}$ is the mass hyperboloid.
In the next subsection~\ref{subsec6.2}, we will argue that, if we choose $\alpha =1$, then the limit exist, and is in fact
independent of $E$, and we have, in the sense of distributional boundary values:
\ben\label{blim}
\lim_{\epsilon \to 0+} \,\, \bv_{\im E < 0, \im E \to 0} B(\epsilon, E,p,s) = (2\omega_p)^{-1} \, {\mathcal B}_p(s) \, ,
\een
plus a contribution that is the boundary value of an analytic function for $\im E > 0$, but that will not contribute to
the expression eq.~\eqref{eq:preBeqn1}. Indeed, for that contribution, we can deform the contour of the $dE$-integration in eq.~\eqref{eq:preBeqn1} to
the trivial contour within the half plane $\im E > 0$, as the exponent $\e^{iE(T-s)}$ provides a damping there
(note that $s<T$), and the same applies to the other $dE_i$-integrals in eq.~\eqref{eq:preBeqn1}. Substituting the limit~\eqref{blim}
into the limit of pre-Boltzmann equation~\eqref{eq:preBeqn1} then delivers the final result
\ben\label{finalmaster}
\begin{split}
\omega_p \partial_T \, \nu_p(T) &=  \B_p(T) \hspace{4mm} - \\
&  \sum_{n=1}^\infty 
\,\,
\Int_{0<\tau_1<...<\tau_n<T}  d^n \tau \,\,
\Int_{{\mathbb H}(m)^n} d\Pi_{k_1,\dots,k_n} \,\, \B_{k_1}(\tau_1)  \frac{\delta}{\delta \nu_{k_1}} \B_{k_2}(\tau_1) \cdots
\frac{\delta}{\delta \nu_{k_n}} \B_{p}(\tau_n)
\, ,
\end{split}
\een
because each $dE$ integration in the first integral in eq.~\eqref{eq:preBeqn1} now yields a delta-function\footnote{
Note that when we substitute eq.~\eqref{blim}, we may take the $ds$-integration from $0$ to $\infty$
when we take the limit, because the integral from $T$ to $\infty$ does not make a contribution
as the $dE$ integration contour can then be deformed  to the trivial contour within the domain $\im E < 0$.},
the effect of which is that the subsequent
$ds$-integration can be performed trivially. The same remark applies to the other iterated integrals on the right side.
This equation is the main result of this paper. It shows how the rescaled number densities $\nu_p$ evolve with time
in the long-time, dilute state limit. Let us discuss the interpretation of eq.~\eqref{finalmaster}.

\medskip
\noindent
We first remark that our equation~\eqref{finalmaster} is different from the Boltzmann equation, see eq.~\eqref{boltzstan},
in that\footnote{There are also trivial differences arising from the relativistic kinematical factors in our equation,
but these are standard and expected, since we are in relativistic model.} the collision factor $\B_p$ involves also collisions other than $2 \to 2$ processes, and in that there are additional ``rescattering terms'' in our equation (the terms with $n>0$), which are non-local in time, and involve iterated collision kernels $\B_p$. These differences disappear if we
additionally consider the case that $\lambda$, the coupling constant, is small. In that case,
all processes with more than 2 incoming or outgoing particles are suppressed, and all
the higher ($n\ge 1$) ``rescattering terms'' in eq.~\eqref{finalmaster} are suppressed by powers of
$\lambda$. Thus, the leading
contribution will arise from the $2 \to 2$ scattering processes and a single collision factor. The corresponding leading approximation of eq.~\eqref{finalmaster} for small $\lambda$ is hence (denoting $\nu_{q_1} = \nu_1, \nu_{q_2} = \nu_2$ etc.):
\begin{equation}\label{eq:Beqn_low_final}
\omega_1 \partial_T \, \nu_1 = 4\pi^2
\Int d\Pi_2 d\Pi_{1'} d\Pi_{2'} \,
\delta^2( q_1^{} + q_2^{} -  q_1' - q_2')
%
\,
| {\mathcal M}_{1, \, 2 \to 1', \, 2'}
|^2 
(
\nu_{1'} \, \nu_{2'}  -  \nu_{1} \,  \nu_{2}
) \, ,
\end{equation}
where the matrix element is now denoting the {\em Born approximation}.
This is indeed the relativistic\footnote{Note the
relativistic kinematical factors implicit in $d\Pi$, as well as in the expression $\omega_p \partial_T$,
which is equal to $p^\mu \partial_\mu$ for the homogeneous state that we consider, because the
$\nu_p$'s are independent of the spatial coordinate.} version of the familiar Boltzmann equation as
given already in the introduction.

In order to get a somewhat better qualitative conceptual understanding when the ``rescattering terms''
can be neglected, let us introduce the $L^1$-norm of a function $f_p$ on the mass hyperboloid ${\mathbb H}(m)$
as $\|f\|_{L^1} = \int |f_p| \, d\Pi_p$. Then the $L^1$-norm of the $n$-th rescattering term in eq.~\eqref{finalmaster}
as a function of $p$ is immediately estimated by
\ben
\begin{split}
&\Int_{0<\tau_1<...<\tau_n<T}  d^n \tau \, \|\B(\tau_1)\|_{L^1} \,
\left\| {\delta \B(\tau_1)}/{ \delta \nu} \right\|_{L^1 \to L^1} \cdots \left\| {\delta \B(\tau_n)}/{ \delta \nu} \right\|_{L^1 \to L^1} \\
& \hspace{2cm} \le \frac{1}{n!} \left( T \sup_{0<s<T}  \sup_{k \in {\mathbb H}(m)} \Int_{{\mathbb H}(m)}
\left| \frac{\delta \B_p(s)}{ \delta \nu_k} \right| \, d\Pi_p \right)^n \equiv
\frac{(T/T_0)^n}{n!} \, .
\end{split}
\een
Here, $T_0$ is defined by the last equation, and $\delta \B(s)/\delta \nu$ is the operator
from $L^1 \to L^1$ that is defined by the kernel $\delta \B_p(s)/\delta \nu_k$.
It is not difficult to see (compare the discussion on p.~93 of~\cite{huang})
that $T_0$ is interpreted as a time of the order of the maximum collision time
(i.e. the average time between two collisions) for the particles of arbitrary momentum $k$
in the medium, between time zero and time $T$. The estimate hence tells us that we
are allowed to drop the rescattering terms if $T \ll T_0$. Now, the physical time
over which the system is observed has actually been rescaled as $t = T/\epsilon$,
by eq.~\eqref{rescalings}, and we have in fact even taken the limit as $\epsilon \to 0$.
Therefore, in terms of the physical time $t$, the condition that
$T \ll T_0$ would mean, for finite but very small $\epsilon$, that $\epsilon t \ll T_0$,
which would appear to be reasonable.

As an aside, we also note that the estimate tells us that
if we could actually mathematically prove that $T_0$ was non-zero, then the series
in eq.~\eqref{finalmaster} would converge. We strongly believe this to be the case,
but have not attempted to prove this. Note however, that in the case of the
pre-Boltzmann, convergence of the corresponding series was proved, and this ought
to provide a good indication here, too.

\subsection{Limit of local S-matrix elements}\label{subsec6.2}

Taking the limit of the collision factor, eq.~\eqref{bepsform}, involves taking the limit as $\epsilon \to 0$ of the matrix elements ${\mathcal M}_X(0,t/\epsilon)$. It is clear from the
perturbative expression for these matrix elements given above in eq.~\eqref{mxypert} that, for finite $\epsilon$,
${\mathcal M}_X(0,t/\epsilon)$ viewed as a function of $X = \{q_1, \dots, q_n\} \in \mr^{2n}$ resp.
$Y = \{q_1', \dots, q_{n'}'\} \in \mr^{2n}$
is an analytic function in the variables $E_i = q_i^0$ and $E_i'= q_i^{\prime 0}$. When $\epsilon \to 0+$, the limit, if it exists,
will still be analytic for $\im E_i < 0$ and $\im E_i' > 0$, because in the integral over the time coordinates
in eq.~\eqref{mxypert}, we can safely continue the frequency arguments in the exponentials to the indicated domain. In other words,
we expect that there exists a function $\F_{X \to Y}$, analytic in this domain such that, in the sense of distributions
\ben\label{bvdef}
\lim_{\epsilon \to 0+} {\mathcal M}_{X \to Y}(0, t/\epsilon) =
\bv_{\substack{
E_i \to \omega(q_i),~ E_i' \to \omega(q_i')\\
\im E_i < 0, \,\,  \im E_i' > 0
}
}
\F_{X \to Y}(\{(E_i, q_i)\}, \{(E_i', q_i')\}) \, .
\een
Here, ``B.V.'' means the distributional boundary value of an analytic function.
The existence of this limit follows from the work of~\cite{egb},
to arbitrary orders in perturbation theory\footnote{Here, it is essential that one takes the parameter $m$ in the
free Hamiltonian to be the true physical mass of the theory. It should also be noted that~\cite{egb} define the adiabatic
limit in terms of some sort of averaging procedure in momentum space around the mass hyperboloids, rather than a
boundary value prescription as above. However, the two are seen to be equivalent.}.
One would also expect this to be true non-perturbatively, but we have not
been able to see this. In the following, we will denote by ${\mathcal M}_{X \to Y}(0, \infty)$ this distributional limit. It corresponds to
the matrix elements of the scattering matrix with an interacting turned on at time $t=0$. The relation to the full matrix element
is easily seen from the following formal calculation:
\ben\label{formcalc}
\begin{split}
& {\mathcal M}_{X \to Y}(-\infty, \infty) \\
&= -\Int_{-\infty}^{+\infty} dt \, \frac{\partial}{\partial t} {\mathcal M}_{X \to Y}(t, \infty) \\
&= i(\omega_X - \omega_Y) \Int_{-\infty}^{+\infty} dt \, \e^{it(\omega_X - \omega_Y)} {\mathcal M}_{X \to Y}(0, \infty) \\
&\vspace{0.4cm}\\
&= 2 \pi \delta(\omega_X - \omega_Y) \bv_{\substack
{
E_i \to \omega(q_i),~ E_i' \to \omega(q_i')\\
\im E_i < 0, \,\,  \im E_i' > 0
}
}
i(E_X - E_Y)
\F_{X \to Y}(\{(E_i, q_i)\}, \{(E_i', q_i')\}) \\
&\vspace{0.4cm}\\
& = 2 \pi \delta(\omega_X-\omega_Y) \left\{ i(\omega_X-\omega_Y) {\mathcal M}_{X \to Y}(0,\infty) \right\}
\, .
\end{split}
\een
In the second line, we have used the identity~\eqref{mxya}, whereas in the last line we have written out the distributional
definition given above in eq.~\eqref{bvdef}. The non-trivial statement is here of course that
the boundary value of $(E_X-E_Y) \F_{X \to Y}$ can indeed be restricted to $\omega_X = \omega_Y$, which is
certainly not obvious, but which can be seen using arguments of~\cite{eg}.

Thus, up to a standard energy-conservation delta-function, the
full matrix element is equal to the scattering matrix element with interaction turned on at time $t=0$, multiplied
by the energy. We also note the distributional equality
\ben\label{distid}
\begin{split}
\frac{\partial}{\partial s} {\mathcal M}(s, \infty) \bigg|_{s=0} &= -i(\omega_X - \omega_Y) \, {\mathcal M}(0, \infty) \, ,\\
&\equiv \bv_{\substack
{
E_i \to \omega(q_i),~ E_i' \to \omega(q_i')\\
\im E_i < 0, \,\,  \im E_i' > 0
}
}
-i(E_X - E_Y) \F_{X \to Y}
\end{split}
\een
which also has been used in the second line.

With this preparation in place, we are now ready to take the $\epsilon \to 0+$ limit of the
collision factors~\eqref{bepsform}. Using eq.~\eqref{mxya}, we can write the relevant integral as
\bena
&&
\epsilon \Int_\mr dt \,\, \e^{-iEt}
E^2 \left| {\mathcal M}_{X \to Y}(0 , t/\epsilon) \right|^2 = \\
&&
2E \Int_0^\infty dt \, \e^{-iEt} \, \re \left\{
\overline{
{\mathcal M}_{X \to Y}(0, t/\epsilon)
} \frac{\partial}{\partial s} {\mathcal M}_{X \to Y}(s, t/\epsilon)
\right\}_{s=0} + \non\\
&&
2E \Int_{-\infty}^0 dt \, \e^{-iEt} \, \re \left\{
\overline{
{\mathcal M}_{X \to Y}(0, t/\epsilon)
} \frac{\partial}{\partial s} {\mathcal M}_{X \to Y}(s, t/\epsilon)
\right\}_{s=0} \non \, .
\eena
The first integral on the right side is analytic in $\im E < 0$, whereas the second in $\im E > 0$.
The latter will hence not make a contribution when we insert the collision factor into the pre-Boltzmann
equation, because we can then deform the contour of the $dE$- (resp. $dE_i$) integrations into the lower complex half-plane
and get zero. Thus, we only need to consider the first integral. The limit as $\epsilon \to 0+$ then renders the
integrand independent of $t$ except for $\e^{-itE}$. So can trivially perform the $dt$-integration which yields simply a factor
of $(E - i0)^{-1}$, and we also use the expressions~\eqref{distid}, and~\eqref{formcalc}. The result is:
\bena
&&
\lim_{\epsilon \to 0+} \epsilon \Int_\mr dt \,\, \e^{-iEt}
E^2 \Big| {\mathcal M}_{X \to Y} (0 , t/\epsilon ) \Big|^2\\
&&
=
4 \pi \Big|
(\omega_X - \omega_Y){\mathcal M}_{X \to Y}(0,\infty)
\Big|^2 \re \left\{
\bv_{\substack
{
E_i \to \omega(q_i),~ E_i' \to \omega(q_i')\\
\im E_i <0, \,\,  \im E_i' > 0
}
}
 \frac{i}{E_X - E_Y} \right\}
\, .
\non
\eena
If we now make use of the distributional identity
\ben
\im \frac{1}{\omega_X - \omega_Y - i0} = \pi \delta(\omega_X - \omega_Y) \, ,
\een
and eq.~\eqref{formcalc}, we arrive at the final expression
\ben\label{mlim}
\lim_{\epsilon \to 0+} \epsilon \Int_\mr dt \,\, \e^{-iEt}
E^2 \Big| {\mathcal M}_{X \to Y} (0 , t/\epsilon ) \Big|^2
= 4\pi^2 \delta(\omega_X - \omega_Y) \Big| {\mathcal M}_{X \to Y} \Big|^2 \, ,
\een
where ${\mathcal M}_{X \to Y}$ is now the full scattering matrix element of the theory,
with the energy-momentum conservation delta's taken out. This expression no longer depends
on $E$.
We use this expression for the limit in order to evaluate the limit
of the collision factors $B(\epsilon, E, p, s)$, see eq.~\eqref{bepsform}. It is clear that
we will only get a finite limit if $\alpha$ and $X,Y$ in that expression are chosen so
that $(|X|-1)\alpha - 1 \ge 0$ and $(|Y|-1)\alpha - 1 \ge 0$. The delta-function $\delta_X(p)$ enforces that $p \in X$.
The energy conservation delta-function $\delta(\omega_X - \omega_Y)$ in eq.~\eqref{mlim} combines with
the momentum conservation delta-function $\delta(k_X-k_Y)$ in eq.~\eqref{bepsform} to an energy-momentum
conservation delta. It then follows that the matrix element ${\mathcal M}_{X \to Y}$ is
zero for $|X| = 1$ incoming particle (or $|Y|=1$ outgoing particle),
because this is kinematically forbidden by energy momentum-conservation\footnote{This can be
different e.g. in a theory containing several particles with different masses $m_i$. In that case,
corresponding changes to the collision factor $\B_p$ would apply.}.
Thus, we must have $|X| \ge 2$ incoming particles in the first term, and then we must choose
$\alpha = 1$ in order to get a finite non-zero limit. It then follows that contributions
with $|X| > 2$ will not contribute. Similarly, in the second term, only
processes with $|Y|=2$ outgoing particles will contribute. Setting $\alpha=1$ in eq.~\eqref{bepsform},
and employing the limit~\eqref{mlim}, we then arrive at the desired limit \eqref{blim} immediately.

\section{Conclusions and outlook}

In this paper, we have demonstrated how the Boltzmann collision equation arises within quantum field
theory. The model that we studied was that of a hermitian, scalar field with polynomial self-interaction.
Employing the projection method and techniques from constructive quantum field theory, we first derived a pre-Boltzmann equation [see eqs.~\eqref{eq:preBeqn},~\eqref{B1}], which was shown to be valid exactly and non-perturbatively in finite volume. This equation has to some extent a similar structure than the
usual Boltzmann equation. On the left side of the pre-Boltzmann equation, we have the time-derivative of the particle number densities $n_k(t)$, whereas on
the right side we have an integral expression involving the particle number densities. The integral expression consists of
iterated collision kernels, which can be expressed in terms of the number densities and scattering matrix elements of the theory,
see eq.~\eqref{B1}. However, there are the following key differences between the pre-Boltzmann equation and the standard
Boltzmann equation:

\begin{enumerate}
\item
The collision terms depend on the {\em local} $S$-matrix elements (i.e. with interaction switched on and off), with
{\em dressed} Feynman propagators [see eq.~\eqref{tildelta}].
\item
There are terms with an arbitrarily large number of collision factors corresponding to ``rescattering''.
\item
The right side of the pre-Boltzmann equation is non-local in time, i.e. depends on the number densities
$n_k(s)$ for $s$ between time $t$ and the initial time.
\end{enumerate}

We then investigated the dilute-medium and long time-limit (and also thermodynamic limit).
Here, we allowed ourselves to interchange certain limits with integrals, and we also made
an assumption about the existence of various limits. In this respect, that part of our analysis
was not entirely rigorous, but we emphasize that we did not drop by hand any terms in the various
expansions considered. In the long-time-dilute-medium limit, we thereby obtained a new, simpler, equation [see eqs.~\eqref{eq:Beqn_low_final0} and~\eqref{finalmaster}]. It differs from the pre-Boltzmann
equation in the following respects:
\begin{enumerate}
\item
The dressed, local matrix elements get replaced by the usual matrix elements (containing all loop orders) with
{\em standard} Feynman propagators.
\item
Scattering processes other than $2 \to n$ particles disappear.
\end{enumerate}

However, there are still multiple rescattering terms. These disappear if we assume additionally that the
coupling constant is small, or that the observation time is shorter than the average time between two collisions. In that case,
we obtain the standard Boltzmann equation, with the scattering matrix elements in the Born-approximation. But if
we want to include higher loop corrections, then the rescattering terms of the corresponding order must also
be taken into account for consistency.

\medskip
\noindent
The following points were left open in this paper:
\begin{enumerate}
\item
We did not justify rigorously the long-time-low-density limit. We believe this to be possible with the machinery
developed for the derivation of the pre-Boltzmann equation, but we
have not attempted to do this. This should be done to confirm the technical correctness of our conclusions with regards
to the importance of the rescattering terms.

\item
In this paper, we have analyzed the long-time-low-density limit, by parameterizing the quantities of
interest in terms of a small parameter $\epsilon$, such that $\epsilon$ is the order of magnitude of the
densities, and such that $\epsilon^{-1}$ is the order of the time duration over which we observe. We have taken
in effect the $\epsilon \to 0$ limit of our pre-Boltzmann
equation. But our framework also allows us to consider a systematic expansion in powers of $\epsilon$, which
would correspond to calculating the sub-leading corrections to the low-density-long-time limit.  We will then encounter,
among other things, propagators with a modified ``dispersion relation'' according to
\ben
E = E(p, \{n_k(t)\}) = \omega_p + O(\epsilon) \, ,
\een
in the Boltzmann equation, where the functional form of $E$ will be determined by a self-consistency requirement.
This is physically reasonable, because it tells us that for a medium that is not dilute, the scattering particles
will feel the effects of the surrounding ``bath'' of particles and can no longer be treated as free, but is typically
not taken into account. We plan to
come back to this issue in a future publication~\cite{hl2}.

\item
It is natural to ask to what extent our formalism can be generalized to a curved spacetime of Robertson-Walker
type $ds^2 = -dt^2 + a(t)^2 dx^2$. In as far as a perturbative analysis (to all orders is concerned), this is possible
using our framework. In essence, we should (i.) employ a version of the projection technique applicable to time-dependent
Hamiltonians, and (ii.) define the ``number densities'' in a way which takes into account particle creation effects in an
expanding universe. (i.) is clearly possible and was already described in sec.~\ref{sec2}. (ii.) is possible if
we replace the ``mode functions'' $u_k$ [see eq.~\eqref{ukt}] by suitable ``adiabatic'' ones involving a WKB-expansion
around flat space, see~\cite{lr,parker}. We will come back to this in a future publication~\cite{hl2}.

We should mention that~\cite{hohen} also claim to have obtained a Boltzmann equation in curved spacetime.
They are working with the so-called Kadanoff-Baym-formalism (see e.g.~\cite{nair} for this approach),
which is closely related also to the Dyson-Schwinger equations.
The end result of their analysis seems to be that the Boltzmann equation remains valid (with the expected changes on
the left side) on a curved manifold, with the flat scattering matrix elements to higher loop order on the right side. Unfortunately, these authors pay insufficient attention to the ambiguities in the definition of number densities in curved space, the absence of a vacuum, and related fundamental difficulties. Also, the nature of their approximations remains rather obscure, and in particular rescattering effects are not taken into account.

\item
One of the major motivations for this work was to understand to what extent it is justified to take into account
loop corrections in the matrix elements in the Boltzmann equation. This question is of considerable importance
e.g. in the context of baryogensis, where baryon-non-conserving processes (violating also $P,CP$) are considered.
The net effect of such processes in the Boltzmann equation is invisible in the Born approximation, and the leading order effect instead comes about through loop corrections. Our analysis indicates that, for consistency, one should then also include corresponding rescattering terms at the corresponding order, but this is normally not done. It would be very interesting to see to what extent this statement affects the analysis of various baryogensis scenarios.

\item
In \cite{buch}, a general formalism was introduced in order to describe a class of states representing some kind
of local thermal equilibrium. Such states are defined by demanding that the expectation values of certain local
operators $W = W[\phi, \dots, \partial^k \phi]$ have the same expectation value within an open spacetime
neighborhood than they would have in a Gibbs-state (possibly with spacetime-dependent temperature). It would
be interesting to see whether our general formalism could also be applied in such situations, and what type of
effective equations might arise.
\end{enumerate}

We believe that these points are worthy of further investigation.

\vspace{1cm}
\noindent
{\bf Acknowledgements:} S.H. would like to thank the Institute of Particle and Nuclear
Studies, KEK, Japan, for its hospitality and financial support during his visit in January
2010, where part of this work was completed. G.L. would like to thank the Slovene Human Resources Development and Scholarship Fund ``Ad Futura'' for a scholarship.



\begin{thebibliography}{99}
\bibitem{ben}
D. Benedetto, F. Castella, R. Esposito, and M. Pulvirenti,
A Short Review on the Derivation of the Nonlinear Quantum Boltzmann Equations,
Commun. Math. Sci. Volume {\bf 5} (2007), 55-71.



\bibitem{b}
N. N. Bogoliubov, D. V. Shirkov, The Theory of Quantized Fields. New York, Interscience,
(1959)

\bibitem{bbgky}
See e.g. N.N. Bogoliubov, Problems of a dynamical theory in statistical physics,
in {\em Studies in Statistical Mechanics}, J. deBoer and G.E. Uhlenbeck eds., Vol {\bf 1},
North Holland, Amsterdam (1962)

\bibitem{bf}
 R.~Brunetti and K.~Fredenhagen,
  Microlocal analysis and interacting quantum field theories:
  Renormalization on physical backgrounds,
  Commun.\ Math.\ Phys.\  {\bf 208} (2000) 623

\bibitem{buch}
 D.~Buchholz, I.~Ojima and H.~Roos,
  Thermodynamic properties of non-equilibrium states in quantum field
  theory,
  Annals Phys.\  {\bf 297}, 219 (2002)


\bibitem{ch}
T. Chen, Boltzmann Limit and Quasifreeness for a Homogeneous Fermi Gas in a Weakly Disordered Random Medium
J. Stat. Phys. {\bf 132} 1572-9613 (2008)

\bibitem{chen1}
T. Chen, Localization lenths and the Boltzmann Limit for the Anderson Model at small
disorders in dimension three, Preprint xxx.lanl.gov/math-ph/0305051

\bibitem{chen2}
T. Chen, $L^r$-convergence of a random Schr\" odinger to a linear Boltzmann Evolution,
Preprint xxx.lanl.gov/math-ph/0407037






\bibitem{df}
 M.~Duetsch and K.~Fredenhagen,
  Causal perturbation theory in terms of retarded products, and a proof  of
  the action Ward identity,
  arXiv:hep-th/0403213.


\bibitem{eng}
D. Eng, L. Erd\" os, The linear Boltzmann equation as the low density limit of
a ransom Schr\" odinger equation, Preprint arXiv:math-ph/0412044 (2005)

\bibitem{eg}
  H.~Epstein and V.~Glaser,
  The Role of locality in perturbation theory,
  Annales Poincare Phys.\ Theor.\  A {\bf 19}, 211 (1973).

\bibitem{egb}
H. Epstein and V. Glaser, Adiabatic limit in perturbation theory, CERN preprint Ref.Th. 1344-CERN
(1971)

\bibitem{salm}
L. Erd\" os, M. Salmhofer, H.-T. Yau,
On the Quantum Boltzmann Equation
J. Stat. Phys. 116 (2004) 367-380

\bibitem{salm1}
L. Erd\" os, M. Salmhofer, H.-T. Yau,
Quantum diffusion of the random Schrödinger evolution in the scaling limit I. The non-recollision diagrams.
math-ph/0512014, Acta Mathematica 200 (2008) 211-277;
Quantum diffusion of the random Schrödinger evolution in the scaling limit II. The recollision diagrams.
math-ph/0512015, Comm. Math. Phys. 271 (2007) 1-53

\bibitem{ey1}
L. Erd\" os and H.T. Yau, Linear Boltzmann equation as the weak coupling limit of the random Schr\" odinger
equation, Comm. Pure Appl. Math. {\bf LIII}, 667 (2000)

\bibitem{ey2}
L. Erd\" os and H.T. Yau, Linear Boltzmann equation as the scaling limit of quantum Lorentz gas,
Adv. in Diff. Eq. and Math. Phys. Contemp. Math. {\bf 217}, 137 (1998)

\bibitem{gj}
J. Glimm and A. Jaffe, Quantum field theory and statistical mechanics, Birkh\" auser, Boston 1985

\bibitem{gj1}
J. Glimm and A. Jaffe, Quantum Physics: A Functional Integral Point of View, Springer 2nd Edition (1987)

\bibitem{glz}
V. Glaser, H. Lehmann and W. Zimmermann, Field operators and retarded functions,
Nuovo Cimen. {\bf 6} 1122 (1957)

\bibitem{haag}
  R.~Haag,
  On quantum field theories,
  Kong.\ Dan.\ Vid.\ Sel.\ Mat.\ Fys.\ Med.\  {\bf 29N12}, 1 (1955)
  [Z.\ Phys.\  {\bf 141}, 217 (1955\ PHMAA,46,376-380.1955)].

\bibitem{hk}
R. Hoegh-Krohn, On the spectrum of the space cutoff $P(\phi)$ Hamiltonian in two spacetime dimensions, Commun. Math. Phys.
{\bf 21} 256-260 (1971)

\bibitem{hohen} A. Hohenegger, A. Kartavtsev, and M. Lindner,
Deriving Boltzmann Equations from Kadanoff-Baym Equations in Curved Spacetime,
arXiv 0807.4551v2 (2009)

\bibitem{hl2}
S. Hollands and G. Leiler, On the derivation of the Boltzmann equation in quantum field theory II: Curved spacetime,
in preparation.

\bibitem{hw1}
 S.~Hollands and R.~M.~Wald,
  Local Wick polynomials and time ordered products of quantum fields in
  curved spacetime,
  Commun.\ Math.\ Phys.\  {\bf 223}, 289 (2001)
\bibitem{hw2}
  S.~Hollands and R.~M.~Wald,
  Existence of local covariant time ordered products of quantum fields in
  curved spacetime,
  Commun.\ Math.\ Phys.\  {\bf 231} (2002) 309

\bibitem{hw3}
  S.~Hollands and R.~M.~Wald,
  On the renormalization group in curved spacetime,
  Commun.\ Math.\ Phys.\  {\bf 237} (2003) 123

\bibitem{huang}
K. Huang, Statistical mechanics, Wiley (1987)

\bibitem{hu}
N.M. Hugenholtz, Derivation of the Boltzmann equation for a Fermi. gas. J. Statist. Phys. 32, 231–254, (1983).

\bibitem{kolb}
E.W. Kolb and M.S. Turner, Grand unified theories and the orgin of the baryon asymmetry, Ann. Rev. Nucl. Part. Sci. {\bf 33},
645-696 (1983)


\bibitem{lr}
 C.~Lueders and J.~E.~Roberts,
Local quasiequivalence and adiabatic vacuum states,
  Commun.\ Math.\ Phys.\  {\bf 134} (1990) 29.

\bibitem{nair}
V. P.~Nair, Quantum Field Theory - A modern perspective
Springer, 2005 (Graduate texts in contemporary physics)

\bibitem{rob}
Original references include:
S. Nakajima, On quantum theory of transport phenomena, Prog. Theor. Phys. {\bf 20} 948 (1958);
B. Robertson, Equations of motion in non-equilibrium statistical mechanics,
Phys. Rev. {\bf 144}, 151 (1966); H. Mori, Transport, collective motion, and Brownian motion,
Prog. Theor. Phys. {\bf 33} 423 (1965)


\bibitem{rs}
M. Reed and B. Simon,
Fourier Analysis, Self-Adjointness (Methods of Modern Mathematical Physics, Vol. 2), Academic Press (1975)


\bibitem{rosen}
L. Rosen, The $(\phi^{2n})_2$ quantum field theory: Higher order estimates, Comm. Pure Appl.
Math. {\bf 24} 417 - 457 (1971)


\bibitem{parker}
  L.~Parker,
  Particle creation in expanding universes,
  Phys.\ Rev.\ Lett.\  {\bf 21}, 562 (1968).

\bibitem{steinmann}
  O.~Steinmann,
  Axiomatic approach to perturbative quantum field theory,
  Annales Poincare Phys.\ Theor.\  {\bf 63}, 399 (1995).


\bibitem{sp1}
H. Spohn, Derivation of the transport equation for electrons
moving through random impurities, J. Stat. Phys. {\bf 17}, 385 (1977)

\bibitem{sp2}
H. Spohn, The Lorentz process converges to a random flight process,
Commun. Math. Phys. {\bf 60}, 277 (1978)

\bibitem{vhove}
L. van Hove, Physica {\bf XXI}, 517-540 (1955)
\end{thebibliography}
\end{document}